\documentclass[12pt]{article}

\pdfoutput=1

\usepackage[top=80pt,bottom=85pt,left=85pt,right=85pt]{geometry}
\usepackage{amssymb}
\usepackage{amsmath}
\usepackage{setspace}
\usepackage{sectsty}
\usepackage{cite}
\usepackage{bbm}
\usepackage[utf8]{inputenc}
\usepackage[T1]{fontenc}
\usepackage{comment}
\usepackage[english]{babel}
\usepackage{afterpage}

\usepackage[usenames]{xcolor}
\usepackage{graphicx,subcaption}
\usepackage{setspace}
\usepackage{float}
\usepackage{booktabs}
\usepackage[debug,pageanchor=false]{hyperref}
\definecolor{link}{rgb}{.8,.15,.1}
\definecolor{pigment}{rgb}{0.36, 0.54, 0.66}
\definecolor{pigment2}{rgb}{0.19, 0.55, 0.91}
\definecolor{pigment3}{rgb}{0.2, 0.2, 0.6}
\definecolor{light-gray}{gray}{0.75}
\hypersetup{colorlinks=true,linkcolor=link,citecolor=link,urlcolor=link,linktocpage}

\usepackage{array,multirow,booktabs,longtable}
\usepackage{mathrsfs}
\usepackage{tikz-cd} 
\usetikzlibrary{backgrounds, arrows,calc,shapes,decorations.pathreplacing, decorations.markings, automata,positioning}
\tikzset{%
  >={Latex[width=2mm,length=2mm]},
            base/.style = {rectangle, rounded corners, draw=black,
                           minimum width=4cm, minimum height=1cm,
                           text centered, font=\sffamily},
  activityStarts/.style = {base, fill=orange!15},
       startstop/.style = {base, fill=orange!15},
    activityRuns/.style = {base, fill=orange!15},
         process/.style = {base, minimum width=2.5cm, fill=orange!15,
                           font=\ttfamily},
}

\newcommand{\red}[1]{}

\tikzset{
        cvertex/.style={circle,draw=black,inner sep=1pt,outer sep=3pt},
        vertex/.style={circle,fill=black,inner sep=1pt,outer sep=3pt},
        star/.style={circle,fill=yellow,inner sep=0.75pt,outer sep=0.75pt},
        tvertex/.style={inner sep=1pt,font=\scriptsize},
        gap/.style={inner sep=0.5pt,fill=white}}

\tikzstyle{mybox} = [draw=black, fill=blue!10, very thick,
    rectangle, rounded corners, inner sep=10pt, inner ysep=20pt]
\tikzstyle{boxtitle} =[fill=blue!50, text=white,rectangle,rounded corners]

\newcommand{\cc}{\mathbb{C}}
\newcommand{\zz}{\mathbb{Z}}
\newcommand{\pp}{\mathbb{P}} 
\newcommand{\Ntwo}{{\mathcal{N}=2}}

\newcommand{\W}{\mathcal{W}}
\newcommand{\T}{\mathcal{T}}

\def\cV{\mathcal{V}}

\DeclareMathOperator{\SU}{SU}
\DeclareMathOperator{\U}{U}
\DeclareMathOperator{\SO}{SO}

\DeclareMathOperator{\im}{im}
\DeclareMathOperator{\coker}{coker}

\DeclareMathOperator{\Spec}{Spec}

\newcommand{\todo}[1]{}
\renewcommand{\todo}[1]{{\color{red} TODO: {#1}}}
\renewcommand{\red}[1]{{\color{red} {#1}}}

\newcommand{\be}{\begin{equation}}  
\newcommand{\ee}{\end{equation}}  
\newcommand{\bea}{\begin{align}}
\newcommand{\eea}{\end{align}}
\newcommand{\bp}{\begin{bmatrix*}[r]}  
\newcommand{\ep}{\end{bmatrix*}}  
\newcommand{\bpp}{\begin{bmatrix}}  
\newcommand{\epp}{\end{bmatrix}}  
\newcommand{\bcd}{\begin{center}
\begin{tikzcd}}
\newcommand{\ecd}{\end{tikzcd} \end{center}}

\makeatletter
\@addtoreset{equation}{section}
\makeatother

\begin{document}


\begin{titlepage}

\begin{center}

\vskip .3in \noindent

{\Large \bf{High electric charges in M-theory \\ from quiver varieties}}

\bigskip\bigskip

Andr\'es Collinucci$^a$, Marco Fazzi$^{b}$, David R. Morrison$^{c}$, and Roberto Valandro$^{d}$ \\

\bigskip


\bigskip
{\footnotesize
 \it

$^a$ Service de Physique Th\'eorique et Math\'ematique, Universit\'e Libre de Bruxelles and \\ International Solvay Institutes, Campus Plaine C.P.~231, B-1050 Bruxelles, Belgium\\
\vspace{.25cm}
$^b$ Department of Physics, Technion, 32000 Haifa, Israel\\
\vspace{.25cm}
$^c$ Departments of Mathematics and Physics, University of California Santa Barbara, \\ Santa Barbara, CA 93106, USA\\
\vspace{.25cm}
$^d$ Dipartimento di Fisica, Universit\`a di Trieste, Strada Costiera 11, I-34151 Trieste, Italy \\
and INFN, Sezione di Trieste, Via Valerio 2, I-34127 Trieste, Italy	
}

\vskip .5cm
{\scriptsize \tt collinucci.phys at gmail.com \hspace{1cm} mfazzi at physics.technion.ac.il \\ drm at math.ucsb.edu \hspace{1cm} roberto.valandro at ts.infn.it}

\vskip 1cm
     	{\bf Abstract }
\vskip .1in
\end{center}

M-theory on a Calabi--Yau threefold admitting a small resolution gives rise to an Abelian vector multiplet and a charged hypermultiplet. We introduce into this picture a procedure to construct threefolds that naturally host matter with electric charges up to six. These are built as families of Du Val ADE surfaces (or ALE spaces), and the possible charges correspond to the Dynkin labels of the adjoint of the ADE algebra. In the case of charge two, we give a new derivation of the answer originally obtained by Curto and Morrison, and explicitly relate this construction to the Morrison--Park geometry.  We also give a procedure for constructing higher-charge cases, which can often be applied to F-theory models.
\noindent

\vfill
\eject

\end{titlepage}


\tableofcontents

\newpage 
\section{Introduction} 
\label{sec:intro}

M-theory on certain singular spaces gives rise to non-Abelian gauge symmetries. More specifically, in Calabi--Yau (CY) geometries with non-isolated singularities that admit \emph{crepant} resolutions (i.e.\ those that do not alter the CY condition), Dynkin diagrams show up naturally as a pattern traced out by families of small intersecting two-spheres in the resolved geometry (for non-isolated holomorphic two-spheres that come in a complex one-dimensional family). When this occurs, postulating the presence of light M2-branes wrapping such so-called \emph{vanishing spheres} gives rise to light degrees of freedom that fill out the root system corresponding to the Dynkin diagram. This leads one to expect the effective field theory to contain a non-Abelian gauge multiplet in the limit where areas of the two-spheres approach zero and the geometry becomes singular \cite{enhanced,witten-phases}.  (For rigid two-spheres, we get light matter multiplets instead, typically
charged under the non-Abelian group.)
This picture has been exploited for two decades in the field of \emph{geometric engineering} in string theory and M-theory \cite{klemm-lerche-mayr-vafa-warner,katz-vafa,bershadsky-johansen-pantev-sadov-vafa,katz-klemm-vafa,katz-mayr-vafa,mayr}, as well as in some closely related F-theory constructions \cite{vafa,morrison-vafa1,morrison-vafa2}. In the latter, a direct interpretation of the singular space is not available in all but the simplest cases. However, two strategies connect F-theory to this phenomenon of non-Abelian degrees of freedom: Either one can take a particular limit of M-theory, such that the duality to type IIB compactified on a circle is manifest, and verify the correspondence. Alternatively, one can analyze the F-theory phenomena directly  by considering which $(p,q)$-strings in the type IIB seven-brane background become light. (An important feature of F-theory is in fact the existence of seven-branes of ``exotic'' types, i.e. beyond the D7-branes and O7-planes of perturbative type IIB.)

Much less clear is what to make of possible Abelian gauge symmetries.\footnote{A second, dual, explanation of the Cartan subgroup for the non-Abelian gauge
symmetry is also available \cite{witten-phases}, via M5-branes wrapping the
surfaces swept out by irreducible deformable two-spheres. This idea
can also be applied to surfaces responsible for an Abelian gauge group.} These can also be related to singularities, but not as directly as in the case of simple Lie algebras. Since the inception of F-theory, it has been known that one way to understand $\U(1)$ gauge groups is by seeking sections (or in some cases, multi-sections) of the elliptic fibration of F-theory \cite{morrison-vafa2,aspinwall-katz-morrison,aspinwall-morrison-U1}. In M-theory, it suffices to think of divisors in the compactification space. One aspect is clear, though. A $\U(1)$ gauge symmetry -- represented by a divisor in the total space of the compactification -- shows its face in the geometry \emph{by means of the (massive or massless) matter charged under it},\footnote{By Poincar\'e duality, in the compact case for each divisor in the (resolved) total space there is an algebraic curve in the total space meeting it precisely once.  Such an algebraic curve represents a matter multiplet of charge one under the corresponding $\U(1)$.} be that in the form of  hypermultiplets or chiral multiplets, depending on the amount of supersymmetry in the effective theory.

A massless charged multiplet can be modeled by a light M2-brane wrapping an algebraic curve in the resolution which shrinks to zero area in the non-resolved fibration.  (If the algebraic curve has no embedded deformations, such a shrinking typically contracts the curve to an isolated singular point, i.e., one of codimension at least three.)  On the other hand, a $\U(1)$ subgroup of the gauge group is understood as the `dimensional reduction' of the supergravity three-form field along a two-form that is Poincar\'e dual to the divisor that intersects the resolution sphere.  The intersection number measures the charge, as is easily seen by considering the flux through a sphere surrounding the intersection point.  

We can also work at the non-compact level, and consider the non-gravitational sector of the theory. The primordial example can be seen by looking at M-theory on a conifold. Now the five-dimensional effective theory still has a hypermultiplet from the M2-brane wrapping the vanishing sphere. The $\U(1)$ vector multiplet, however, is no longer localized to five dimensions, but is seven-dimensional.\footnote{To see this, put M-theory on an $A_1$ surface singularity, seen as a limit of a two-center Taub--NUT space. There are two homologically nontrivial two-cycles, call them $N_i$, which are Poincar\'e dual to two harmonic, normalizable two-forms $\omega^i$ (i.e. $\int_{N_i}\omega^j =\delta_i^j$). We can decompose the supergravity $C_3$-form along them as follows: $C_3 \sim \omega^i \wedge A_i$. The $A_i$'s are $\U(1)$ gauge potentials supported on the seven-dimensional external space, and localized on the $i$-th center of the Taub--NUT (that is, the kinetic term $G_4 \wedge G_4$ peaks around the centers) \cite{gubser-tasi}. Now take M-theory on the conifold $uv=zw$, regarded as $\cc^*$-fibration over $\cc^2_{(z,w)}$. The reduction to type IIA gives rise to two intersecting D6-branes, localized at $z=0$ and $w=0$. The two worldvolume $\U(1)$'s live in seven dimensions (and one of them is decoupled \cite{gubser-tasi}), while at their five-dimensional intersection lives a charged hyper.} So it shows up as a flavor symmetry under which the hyper is charged. Throughout this paper, we will work in this non-compact setting. However, it should always be borne in mind that our models can be embedded into full-fledged compact geometries, where the $\U(1)$ will be gauged again. Hence, we will refer to our $\U(1)$'s as gauge symmetries, and consider the hypers as electrically charged, albeit as subsectors of the full theory.

As we stated above, $\U(1)$'s are visible in the geometry through the matter which is charged under them. One could now ask: Given a $\U(1)$ gauge symmetry and an electric charge quantization condition, what are the possible charges available? In perturbative intersecting D-brane models, one can easily find charge-one spectra. By introducing orientifolds, one can also find charges higher than one (see e.g. \cite{Cianci:2018vwv} for recent constructions with charges up to six). However, no explicit F-theory geometries have been constructed that go beyond charge four. (The first charge-three and four models have appeared in \cite{klevers-mayorga-oehlmann-piragua-reuter} and \cite{raghuram} respectively.\footnote{The most general F-theory geometry admitting charge-two matter was constructed in \cite{morrison-park,tall-sections}, where the above question was also raised.} The conditions satisfied by an elliptic fibration realizing a global $\SU(N)$ model with matter multiplets of high $\U(1)$ charge were earlier determined in \cite{Mayrhofer:2012zy,Lawrie:2015hia}. For global models with $\SU(5)$ matter of high $\U(1)$ charge see e.g. \cite{Mayrhofer:2014opa,Mayrhofer:2014haa}.)

This subject has recently been investigated in the F-theory framework in \cite{taylor-turner,raghuram-taylor},\footnote{For pedagogical reviews on F-theory and its recent developments, see \cite{Weigand:2018rez,cvetic-lin-tasi}.  We follow the perspective advocated in \cite{whatF}, in which F-theory is regarded as a variant of type IIB string theory in which the data of the varying elliptic curve is used to specify the variable axio-dilaton of IIB.} where the primary method for constructing higher-charge matter is an indirect one: One starts out with a setup that gives rise to a non-Abelian gauge symmetry with matter in either the adjoint or more ``exotic'' representations (such as the three-index antisymmetric of an $\SU$ group \cite{klevers-taylor,klevers-morrison-raghuram-taylor}). Higgsing  in particular ways one shows that these representations must decompose into matter of higher $\U(1)$ charge.

In this paper, we explore a completely different method that can readily generate a large class of examples of CY threefolds with matter of electric charge up to $q = 6$. Here, the $6$ is the highest possible Dynkin label of any adjoint representation for a simple Lie algebra. In this case, it is an $E_8$ label. We will explain the method in what follows.

The rough idea is to look at families of local surfaces.  Exactly how to ``localize'' surfaces leads to some technical questions, which traditionally have been given different answers in algebraic geometry, complex analytic geometry, and hyper-K\"ahler
geometry.  To fix ideas, our basic building blocks will be asymptotically locally Euclidean (ALE) spaces, which come equipped with hyper-K\"ahler metrics and which can easily be related to the so-called ``rational double point'' singularities in algebraic geometry \cite{kronheimer-ALE}.  Moreover, the deformation
theory can be described in terms of period integrals, which would also be
the case if the ALE spaces were embedded as local surfaces within global
(possibly singular) K3 surfaces.

We will thus  consider families of ALE spaces, often
modeled by algebraic families of rational double point singularities.
Within these families, one can impose that the ALE spaces degenerate in particular ways, so as to generate point-like singularities that admit small resolutions, akin to the conifold. However, unlike the conifold, the non-compact divisor will intersect the exceptional $\pp^1$ $\ell$ times, where $0 \leq \ell \leq 6$, depending on the situation.
In order to construct these families, we will be guided by the principles of Brieskorn--Grothendieck simultaneous resolution \cite{brieskorn-sim,slodowy,slodowy-fourlec} (made explicit in \cite{katz-morrison}), as well as Koll\'ar's notion of higher-length flops \cite[pp.~95-96]{[CKM]}, further developed in \cite{katz-morrison} and \cite{curto-morrison}. Concretely, we will use some technology that has recently been developed in \cite{karmazyn}, whereby one describes not the geometry directly, but a quiver that encompasses all relevant information. Such a quiver, known as the \emph{universal flopping algebra}, has the property that its space of representations is literally a model for the geometry of interest (up to taking completions of local rings), with deformations and resolutions being regarded as Fayet--Iliopoulos terms. The advantage of using such a language is that it gives us a very compact way to package entire families of ALE spaces.

The representation spaces of these quivers actually give us CY $n$-folds, where $n\geq 3$. However, an arbitrary three-dimensional slice (or an appropriate covering of such a slice) will give us a CY threefold that admits a small resolution. By analyzing which homology spheres get contracted along various loci of the CY, we will write down the criteria that such a threefold must satisfy in order for it to admit higher-charged matter.

\subsection*{Summary and outline}
\begin{figure}[ht!]
\centering
\includegraphics[scale=.95]{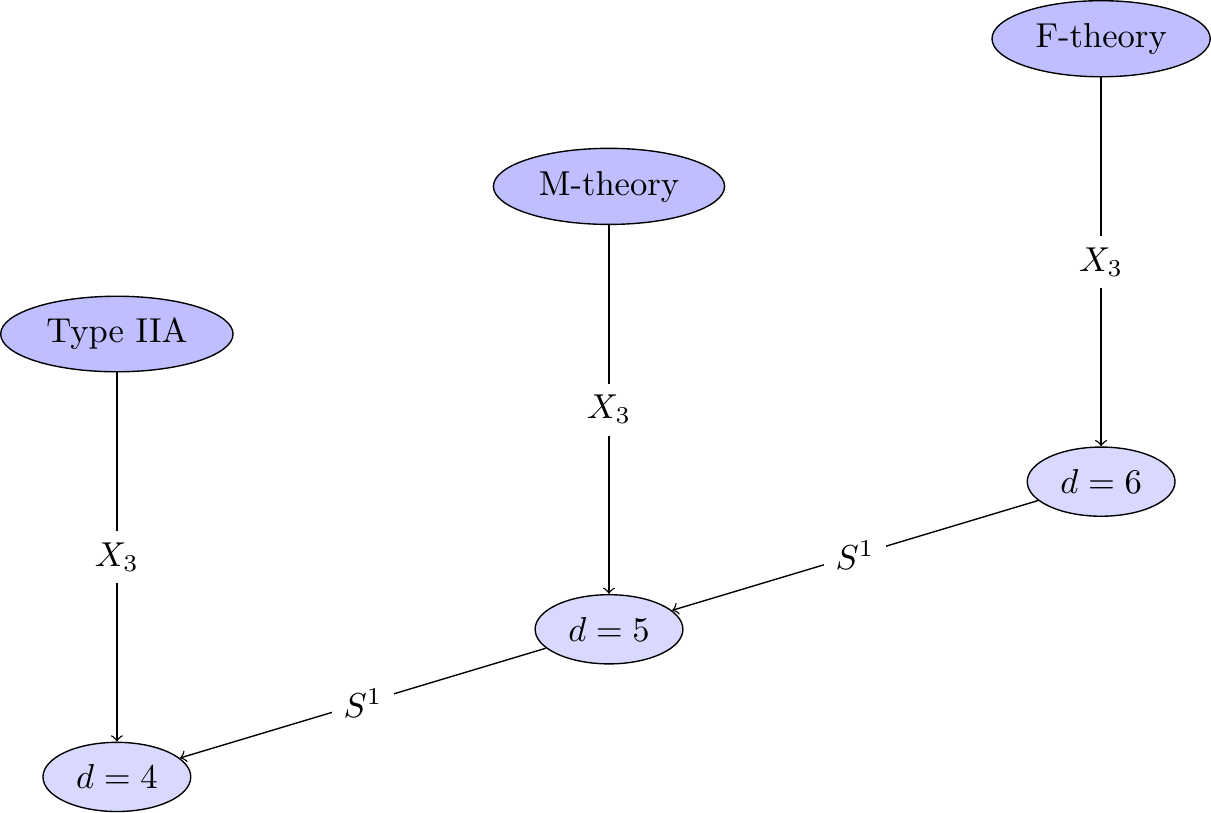}
\caption{The various theories under consideration and their interconnections. (In the case of F-theory, the data of the fibration $X_3\to B_2$ -- when it exists -- is used to specify the type IIB axio-dilaton.)  Each lower-dimensional effective theory has eight Poincaré supercharges.}
\label{fig:theories}
\end{figure}

We will now summarize the main point of our analysis, and give an outline of the sections. However, we would first like to clarify our setup, as names will often get used interchangeably. We will always be considering a non-compact Calabi--Yau threefold $X_3$ that 
admits a simple small resolution, and we will sometimes assume
that $X_3$ can be regarded as a 
local
patch of an elliptic fibration over a base $B_2$ which is an open
subset of $\cc^2$.\footnote{Our
basic assumption is that $X_3$ can be fibered by ALE spaces, but sometimes
we want an elliptic fibration as well so in that case $X_3$ should
be fibered by ALG spaces.} The various possible effective theories obtained via ``compactification'' on $X_3$ are the following:
\begin{itemize}
\item Type IIA on $X_3$, giving four-dimensional $\Ntwo$ supersymmetry.
\item M-theory on $X_3$, giving five-dimensional $\mathcal{N}=1$ supersymmetry.
\item F-theory on $X_3$ with type IIB axio-dilaton specified by the fibration $X_3\to B_2$ (when that fibration exists). 
This gives a six-dimensional theory with $\mathcal{N}=(1,0)$ supersymmetry. 
\end{itemize}
In the compact setting, all three theories have a single $\U(1)$ vector multiplet (and no non-Abelian ones) plus charged hypers, whereby the charges will vary according to our constructions. The interconnections are displayed in figure~\ref{fig:theories}.

\begin{figure}[ht!]
\centering
\vskip 5pt
\includegraphics[scale=1.1]{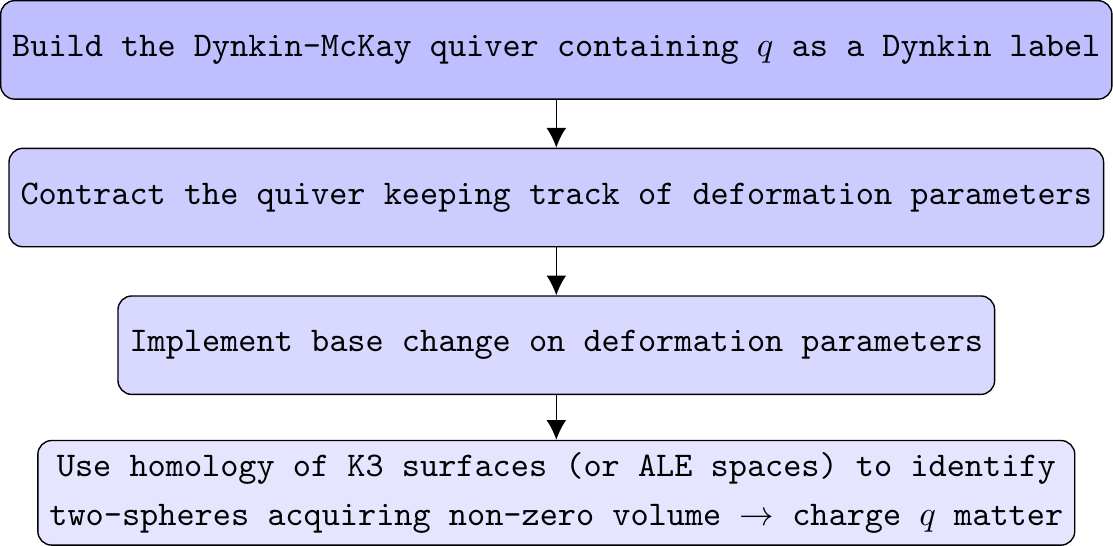}
\caption{The main idea}
\label{fig:summary}
\end{figure}

Let us now summarize our findings. The construction by Karmazyn \cite{karmazyn} of the so-called \emph{universal flopping algebra of length $\ell$} provides us a CY $n$-fold,\footnote{At times, following \cite{curto-morrison}, we will refer to this CY as \emph{the universal flop of length} $\ell$.} from which we can take any three-dimensional slice (or appropriate covering of such a slice). Any such slice or cover will contain a $\U(1)$ vector multiplet, and matter with charge $\ell$.  We will show how this works explicitly for charge two, and relate the construction of length two directly to the well-known Morrison--Park geometry \cite{morrison-park}.\footnote{Note that in the length 2 case, the elliptic fibration and presumably the ALG spaces are a natural feature of the solution.} We will then explain how to obtain higher charges from higher-length flopping algebras. The recipe to obtain charge $q$ is summarized in figure \ref{fig:summary}, whereas the necessary terminology will be explained in later sections.

This paper is organized as follows. In section \ref{sec:families}, we first present the concept of building threefolds as families of surfaces, by demanding that these admit \emph{simultaneous resolutions}. We first whet the reader's appetite with a familiar example, the conifold \cite{stromcon,bhole}, seen as a family of $A_1$ surfaces. We then explain the general picture. In section \ref{sec:quivers}, we specialize to the case of the universal flop of length two (a CY sixfold). This is the case that will ultimately produce threefolds with matter of charge two, starting from the $D_4$ surface. Here, we expose the technology recently developed in \cite{karmazyn}, where one constructs these families in terms of quivers. In section \ref{sec:D4-quiv}, we actually proceed to the construction of threefolds with charge-two matter.\footnote{Our presentation gives a new derivation of results originally obtained in \cite{curto-morrison}, which also formulated a conjectural extension to higher-length cases.} Here, we discover that these threefolds exhibit a rich matter structure. Depending on how we cut the sixfold down to a CY threefold, we may get:
\begin{itemize}
\item Charge-one matter loci, along which the exceptional locus of the (simple small) resolution is a $\pp^1$.
\item We may get loci with \emph{length-two} flops, corresponding to charge-two matter made from a bound state of membranes on the exceptional $\pp^1$. (We give an explicit example of this in section \ref{sub:laufer}.)
\item We may get an exceptional locus consisting in a union of two $\pp^1$'s.
\item Finally, we may get a quadratically embedded $\pp^1$, such that the standard $\U(1)$ divisor cuts it at two points. This produces charge-two matter.
\end{itemize}
The various possibilities are summarized in figure \ref{fig:P1s}. In section \ref{sec:MP}, we establish the correspondence of this construction with the Morrison--Park one.  This is the classic example of F-theory with a single $\U(1)$, and matter hypermultiplets of charge one and two. In section \ref{sec:gen-K3}, we open up the investigation for higher charge, and we list the universal quivers of \cite{karmazyn}. We briefly present our conclusions in section~\ref{sec:conc}. Finally, in appendix \ref{app:explicit} we explain how to construct general threefolds admitting high-charge matter loci with the aid of simple computer algebra. The outputs of this calculation are written in the \texttt{Mathematica} notebook included with the \texttt{arXiv} submission,\footnote{Which can also be found at this \href{http://web.math.ucsb.edu/~drm/mathematica/}{\nolinkurl{page}}.} where we present the most general $n$-folds in elliptic form admitting charge three, four, five, and six.


\subsection{Local surfaces}
Before ending this introduction, we wish to make our conventions about
local surfaces completely explicit, as well as to introduce the important
notions of local and global Picard groups.

In algebraic geometry, the key object is the ring of algebraic functions.
The natural domains of definition of algebraic functions are the so-called
Zariski open sets, and passing to a smaller Zariski open set will allow
algebraic functions with poles along the algebraic subset which was
removed, i.e. the ring of algebraic functions will increase.  The natural
way to ``shrink'' a surface in that case would be to pass to a Zariski open
subset.  Unfortunately, such subsets tend to be rather large.
The ``solution'' to the problem of defining functions which are appropriate 
for even smaller
neighborhoods, is to use functions expressed as formal power series 
(i.e., we should complete the local ring at its maximal ideal).

In complex analytic geometry, the key object is the ring of holomorphic
functions.  The natural domains of definition are much smaller than in the
algebraic case, and various
rings of functions can be related by the ``germ'' construction:  Two functions
are in the same germ at a point $P$ if they agree in a small neighborhood
of $P$.  The ring of germs of functions at $P$ tends to be a bit smaller
than the formal power series ring, even if the space on which the
functions are defined is an algebraic variety, because the 
power series corresponding to holomorphic functions are necessarily convergent.

If we have a singular algebraic variety or a singular complex analytic
space, there is a well-developed local deformation theory which relies
on the rings of functions described above.  The results either describe
a deformation up to (formal) completion, or up to passing to germs of 
functions.

The simplest singularities in complex codimension two, 
which have many equivalent descriptions \cite{Durfee}, 
are the so-called \emph{rational double
points}.  A convenient description of these is as quotients: They all take the
form $\cc^2/\Gamma$ for $\Gamma$ a finite subgroup of $\SU(2)$.
There are many other descriptions.

Kronheimer gave a quotient construction for such singularities and their
deformations, which incidentally puts a metric on the underlying complex
space (even when that space is singular -- in which case the metric is
well-defined on the smooth locus and has controlled asymptotic behavior).
Kronheimer's construction \cite{kronheimer-ALE} puts a hyper-K\"ahler
metric on each fiber.  The singular spaces and their resolutions are
all described as spaces which asymptote (metrically) to $S^3/\Gamma$
at infinite distance -- such spaces (whose metric satisfies an appropriate
equation) are known as asymptotically locally Euclidean, or ALE, spaces.

Kronheimer also describes the deformation theory in terms of the
period integrals of two-forms from the hyper-K\"ahler structure with
respect to compact two-cycles in the space \cite{kronheimer-torelli}.
Many such deformations can be seen as arising from degenerations of
(hyper-K\"ahler) K3 surfaces.  The addition of orbifold metrics
to complete the moduli space of K3 surfaces (the need for which had
been pointed out in \cite{remarksK3}) was demonstrated in
\cite{MR902574} as an extension of Yau's proof of the Calabi conjecture
\cite{MR480350}.  Subsequently,
Anderson \cite{MR1143663} showed that all degenerations of K3 surfaces
which are at finite distance from the bulk of the moduli space can be
realized in this form.  In particular, K3 surfaces may develop orbifold
singularities in the way predicted by Kronheimer's analysis.  Thus, most of
the singularities we are studying could also be modeled on open subsets
of singular K3 surfaces.  (Note however that such a model has a bounded rank -- giving a bound on the number of independent rational curves which can 
shrink -- which is why not all of the singularities of interest can be modeled on
open subsets of K3 surfaces).

In addition to the ALE spaces, geometers and physicists have studied other types of `gravitational instantons', namely ALF (asymptotically locally flat), ALG, and ALH spaces \cite{Hawking:1976jb,MR2869021,arXiv:1505.01790,arXiv:1508.07908,arXiv:1603.08465}, which are all asymptotic to fibrations over a Euclidean base and contain tori (the fibers) of various dimensions at infinity (and
in fact throughout the interior of the space).  Relevant to F-theory are
the ALG spaces which have a $T^2$ fibration;\footnote{For ALF spaces the fibers are circles, whereas for ALH spaces they are compact orientable flat three-manifolds. See e.g. \cite{hein-talk}.} in many instances, the deformation
theory of the ADE singularities can be reproduced by ALG spaces as well as by
ALE spaces.

\subsection{Local and global Picard groups}

In this paper, we measure the charge of various matter fields (represented
by compact curves in the space) with respect to various gauge fields
(represented by possibly non-compact divisors in the space) in terms
of the intersection number between divisors and curves.  In order to compare
charges (and therefore assert that we have matter with high $\U(1)$ charge), 
we need a way to compare divisors at different points of
the space.  This is provided by the theory of the Picard group.

On any algebraic variety or complex analytic space, one can define
the Picard group as the free Abelian group of subvarieties of complex
codimension one modulo the relations obtained by considering the
divisors of zeros and poles of arbitrary meromorphic functions on
the space.  If $D$ and $D'$ are equivalent under such a relation,
i.e.. if $D-D'$ represents the divisor of zeros and poles of some
meromorphic function $f$, then for any compact curve in the space
$D\cdot C = D'\cdot C$.  Thus, for computing gauge charges, one
can pass naturally to the Picard group.  Similarly, the existence
of the meromorphic function $f$ shows that the gauge field
obtained by reducing the M-theory three-form on the divisor $D$ is
gauge equivalent to the gauge field obtained by reducing the three-form
on $D'$.  The Picard group thus captures an Abelian
subgroup of the full gauge group, in which the (Abelian)
gauge fields come by
reduction of the M-theory three-form on divisors.  

The computations we will make of gauge charges in this paper will
typically be local ones:  we will find a small resolution of a  
local neighborhood of some 
singular point $P$, and then consider a compact curve $C_P$ in that small
resolution.  The gauge charge of $C_P$
is then an integer-valued function on the
local Picard group, i.e. the Picard group of divisors defined in
an appropriate neighborhood of $P$.  In principle, the gauge charge
might depend on the choice of small resolution, but such dependencies
can be made very explicit.

A bit more challenging is the problem of relating the gauge charges of 
a curve $C_P$ lying over one singular point $P$, with those of a curve $C_{P'}$
lying over a different singular point $P'$.  In fact, this problem
goes back to the earliest examples of conifold transitions \cite{Candelas:1987kf,bhole,confinement} and their mathematical explanations \cite{clemens1983double,MR848512}.
In the case of a quintic threefold specializing to a quintic $X$ containing
a $\pp^2$, sixteen singular points are created but there is only one
new divisor class (the proper transform of the $\pp^2$)
after performing a small resolution $Y \dashrightarrow X$.  However, there are maps
$\text{Pic}(Y)\to \text{Pic}(Y_P)$ where 
$Y_P$ is the inverse image of a small neighborhood
$U_P$ of the singular point $P$.  In fact, $Y_P$ is just a neighborhood
of a single compact curve $C_P \cong \pp^1$, and one 
shows that the new divisor class
maps to a generator of $\text{Pic}(Y_P)\cong \zz$ 
for each $P$.  In fact, each curve $C_P$ has charge $1$ under
the new divisor class (or perhaps each curve $C_P$ has charge $-1$,
which happens if the small resolution is changed).

\begin{figure}[ht!]
\centering
\includegraphics[scale=.4]{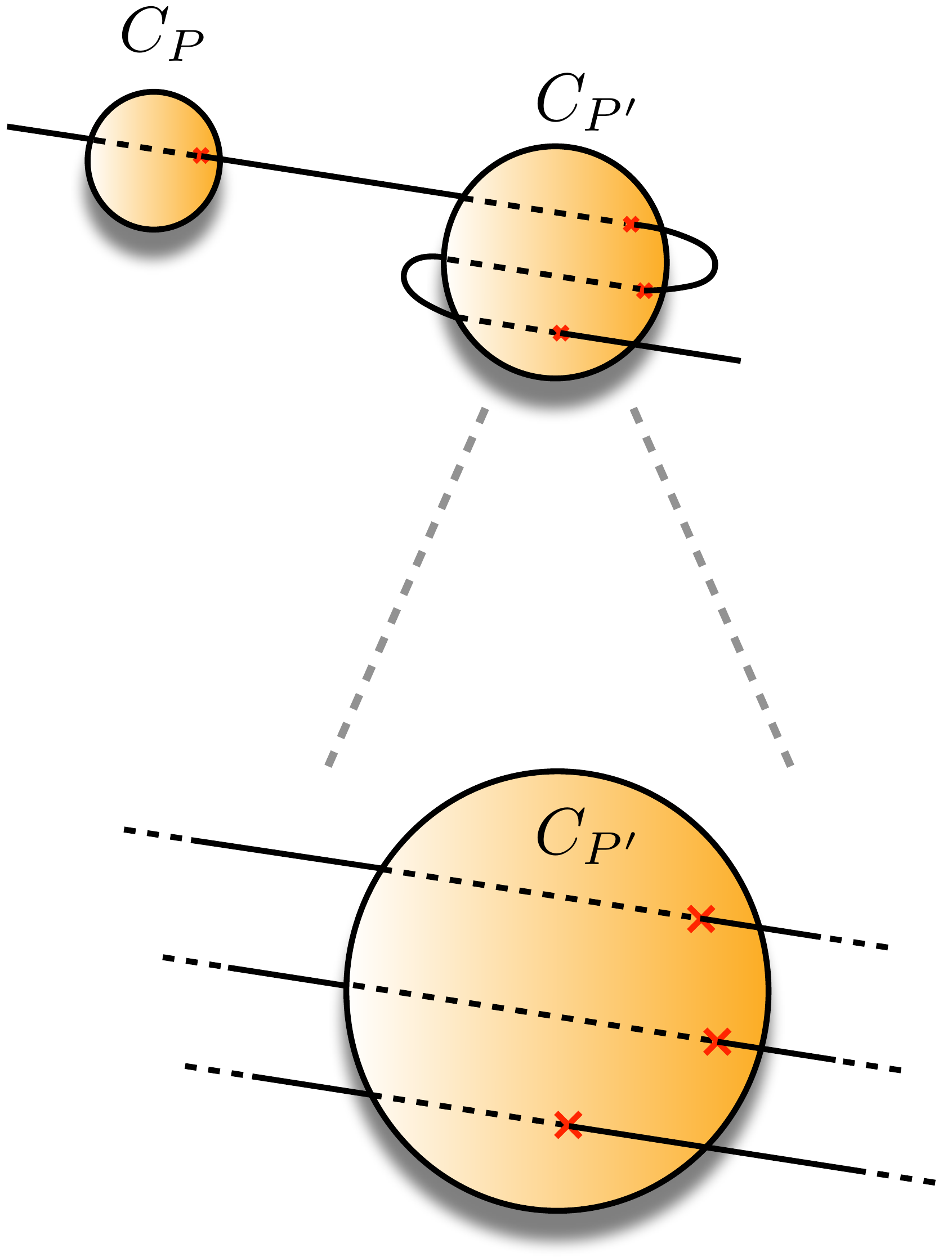}
\caption{A globally defined generator of $\text{Pic}(Y)$ can intersect the resolution curve $C_P$ once, and $C_{P'}$ $N$ times (top frame). However, in a local patch around the curve $C_{P'}$, this looks like $N$ times the generator of $\text{Pic}(Y_{P'})$ (bottom frame).}
\label{fig:picard}
\end{figure}

The basic principle that calculations can be done in local Picard groups
$\text{Pic}(Y_P)$ and then related to each other in the global
Picard group $\text{Pic}(Y)$ is something we will encounter frequently in this paper. Suppose our threefold $Y$ contains two singular points $P$ and $P'$ supporting charged matter. After resolving both singularities and creating exceptional curves $C_{P}$ and $C_{P'}$, we can define the local Picard groups $\text{Pic}(Y_{P})$ and $\text{Pic}(Y_{P'})$. Each local Picard group will contain a divisor that intersects the local exceptional curve once. However, for the full `global' model,\footnote{By global we do \emph{not} necessarily mean compact.} a generator of $\text{Pic}(Y)$ might intersect, say $C_{P}$ once, and $C_{P'}$ $N$ times, where $N$ could be arbitrarily high. See figure \ref{fig:picard} for a cartoon.

This is how higher-charge models are possible. In our paper, we will study a class of models that automatically generate charges up to six. However, even higher charges could be possible if one glues several such models appropriately. This is an interesting direction for future research.

\section{Threefolds as families of ADE surfaces}
\label{sec:families}
\subsection{\texorpdfstring{Elementary example: The conifold as a family of $A_1$ surfaces}{Elementary example: The conifold as a family of A1 surfaces}} 
\label{subsec:A1}

\begin{figure}[ht!]
\centering
\includegraphics[scale=1]{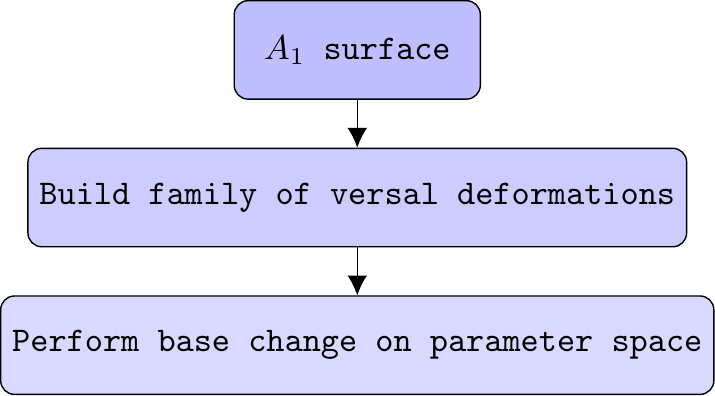}
\caption{The main idea exemplified in the $A_1$ case.}
\label{fig:A1work}
\end{figure}

\begin{figure}[ht!]
\centering
\includegraphics[scale=.5]{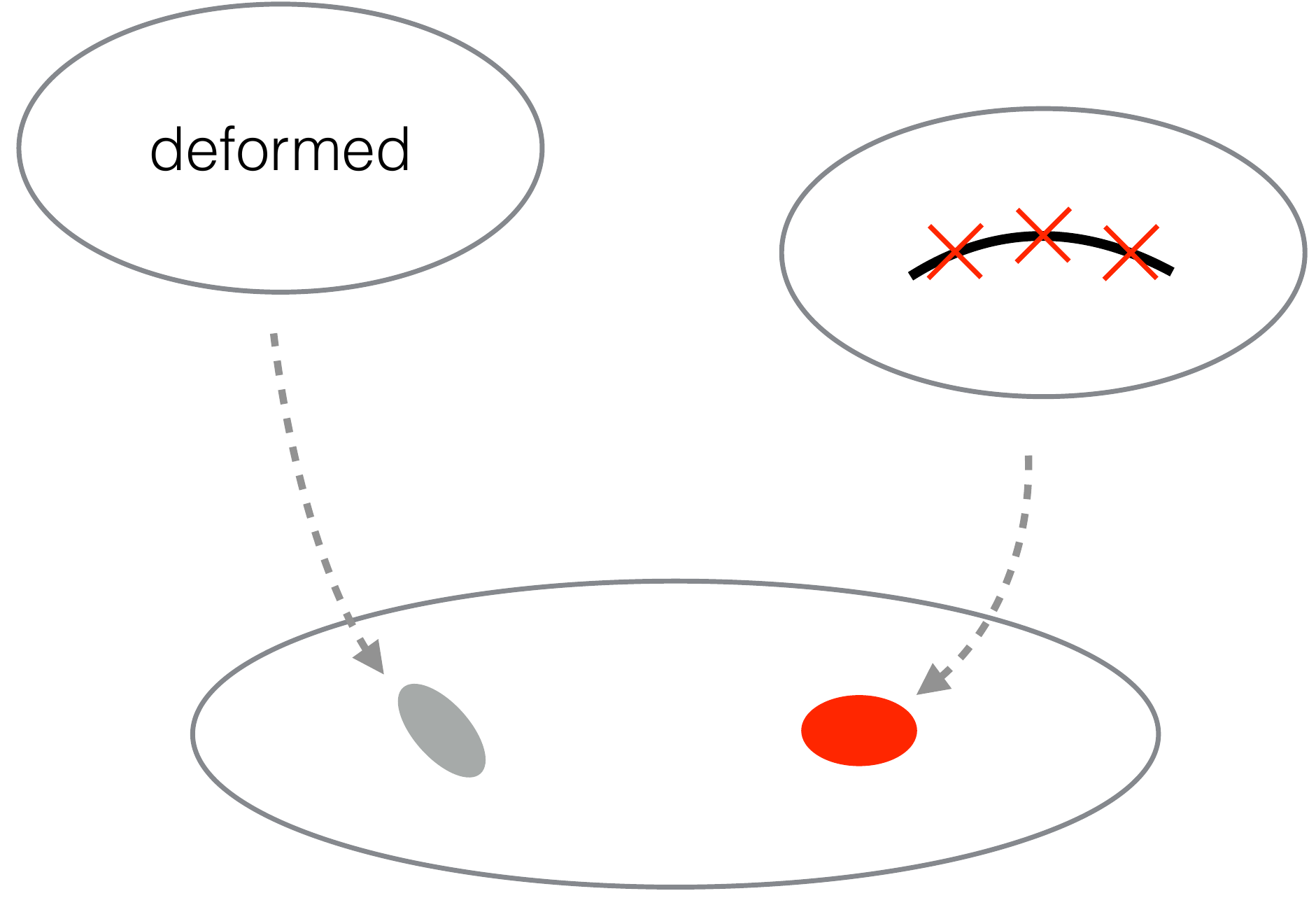}
\caption{A simultaneous resolution of singularities ($D_4$ in the picture). In the `resolved' fiber over the red base locus, a black line represents an exceptional $\pp^1$ that has been blown up, with three marked points (the red crosses) representing residual $A_1$ singularities. Over the generic grey locus, the fiber is completely smooth, and corresponds to a versally deformed surface singularity.}
\label{fig:simultaneous}
\end{figure}

In this section, we will rediscover the simplest of flops, based on the conifold, as a one-parameter family of surfaces.\footnote{The idea of building threefolds by starting with surfaces is not new to physics, as exemplified in \cite{klebanov-witten,nonspherI}. It is also not new to geometry \cite{atiyah,reid}.  We will take a purely geometric stance here.} This gives us a prototype local model for a theory with a $\U(1)$ symmetry and one charged hyper. The basic idea is summarized by the workflow \ref{fig:A1work}.
In equations, we start with the standard Du Val $A_1$ surface given by the following hypersurface:
\begin{equation}
P:=x^2+y^2+z^2=0\ .
\end{equation}
Its family of \emph{versal} deformations is parametrized by the Jacobian ring
\begin{equation}
\mathbb{C}[x, y, z]/(\nabla P) = \mathbb{C}[x, y, z]/(x, y, z) \cong \mathbb{C}\ . 
\end{equation}
In other words, we have the following one-parameter family:
\begin{equation}\label{eq:A1def}
x^2+y^2+z^2=\alpha\ .
\end{equation}
The total space of this family is a threefold $X_3$, and its central fiber over $\alpha = 0$ admits a resolution. However, there is no `simultaneous resolution' for the whole family.\footnote{The question of which deformations of surface singularities admit `simultaneous resolutions' was investigated thoroughly
by Grothendieck \cite{unpublished} and Brieskorn \cite{brieskorn-sim}; see \cite{slodowy} for a detailed exposition.} In other words, we can only perform the birational transformation on the central fiber. A simultaneous resolution would correspond to an operation we would perform on this whole threefold, such that its restriction to the central fiber would be the standard resolution, as schematically presented in figure \ref{fig:simultaneous}.
\begin{figure}[hb!]
\centering
\includegraphics[scale=.4]{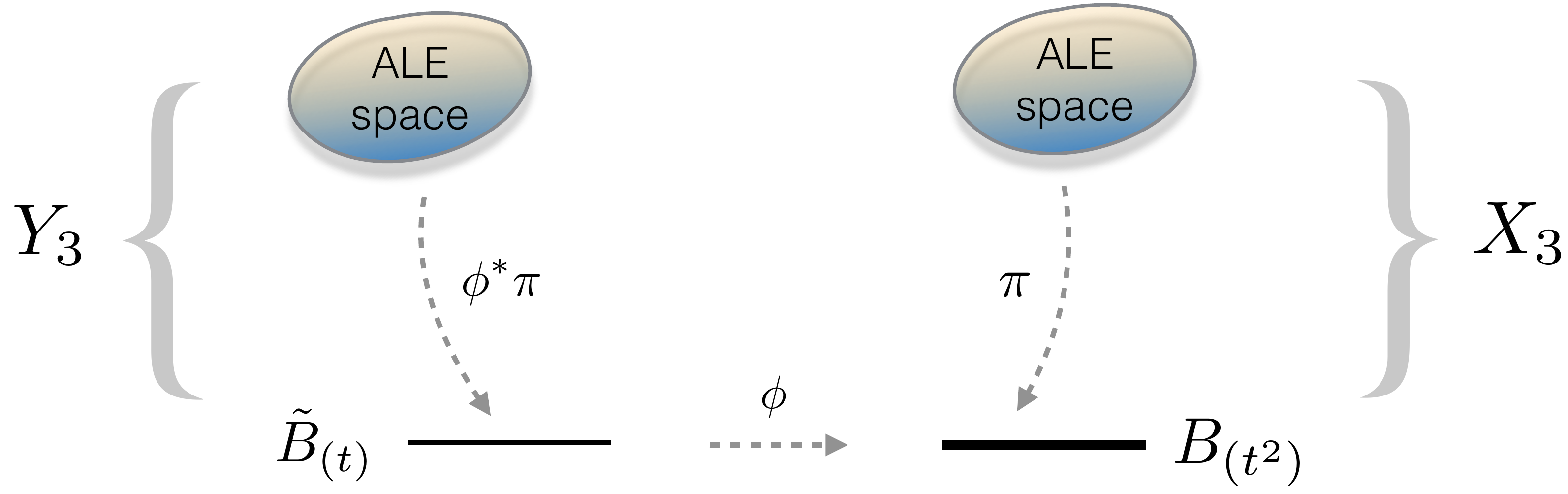}
\caption{The new threefold family $Y_3$ (given as a fibration $\phi^* \pi : \text{ALE} \to \tilde{B}$) after base change.}
\label{fig:basechange}
\end{figure}
Instead, Atiyah observed \cite{atiyah} that if we perform a \emph{base change}, meaning, if we pull back this family onto a particular covering space of the parameter space of $\alpha$, then we will have a new threefold $Y_3$ that does admit a simultaneous resolution. In this case, the base change is (see figure \ref{fig:basechange})
\begin{equation}
\phi:\tilde B \longrightarrow B :\ t \mapsto \alpha  = t^2\ .
\end{equation}
Now, the end result after pulling back the family of surfaces onto $\tilde B$ is the conifold:
\begin{equation}\label{eq:A1coni}
x^2+y^2+z^2 = t^2\,.
\end{equation}
As is well known \cite{bhole}, the conifold admits two `small resolutions', and each one restricts to the usual resolution for the $A_1$ singularity, when we set $t=0$.

\subsection{General ALE surfaces}
\label{subsec:adesurf}
This innocent trick of replacing the deformation parameter $\alpha$ with the square of a new parameter $t$ has underpinnings in the \emph{Weyl invariant theory} of $A_1$. For the case of the conifold, this machinery constitutes overkill. However, it is crucial in the construction of other threefolds that admit simultaneous resolutions \cite{katz-morrison}.

Here is a preview. The Lie group $A_1$ has a one-dimensional Cartan torus $\T$ that we can parametrize with the coordinate $t$. There is a Weyl symmetry group $\W=S_2$ which acts as $t \mapsto -t$. The standard versal deformation of an ALE hypersurface is a family of surfaces parametrized by coordinates of the quotient space $T/\W$ \cite{slodowy-fourlec,slodowy}. In this case, this coordinate is $\alpha$, and the family is 
\begin{equation}
x^2+y^2+z^2=\alpha\ ,
\end{equation}
which does not admit a simultaneous resolution. However, if we pull this family back onto $\T$, and express it in terms of the Weyl-\emph{covariant} coordinate $t=\sqrt{\alpha}$, then we get the conifold, which admits a simultaneous resolution.

The story for more general ALE groups is similar:  there is a finite base change to the Weyl-covariant coordinates which allows for a complete simultaneous resolution.  There are, however, also intermediate cases in which one wishes to simultaneously resolve only along a subgraph of the Dynkin diagram (which can be colored black to aid in visualization). There is then a subgroup $\W'$ of the full Weyl group corresponding to the complement of the subgraph (which would be colored white),
and a partial quotient $\T/\W'$; pulling back the deformation to that partial quotient allows a partial simultaneous resolution.  The intermediate quotient lies between the Cartan torus $\T$ and the full quotient $\T/\W$:
\begin{equation}
\T \rightarrow \T/\W'  \rightarrow \T/\W\ .
\end{equation}
\begin{figure}[ht!]
\centering
\includegraphics[scale=.56]{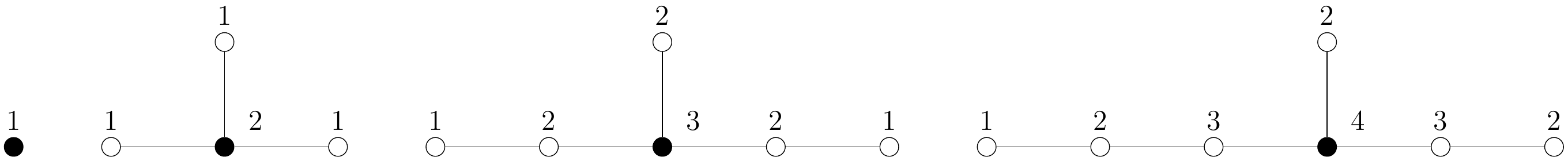}
\includegraphics[scale=.56]{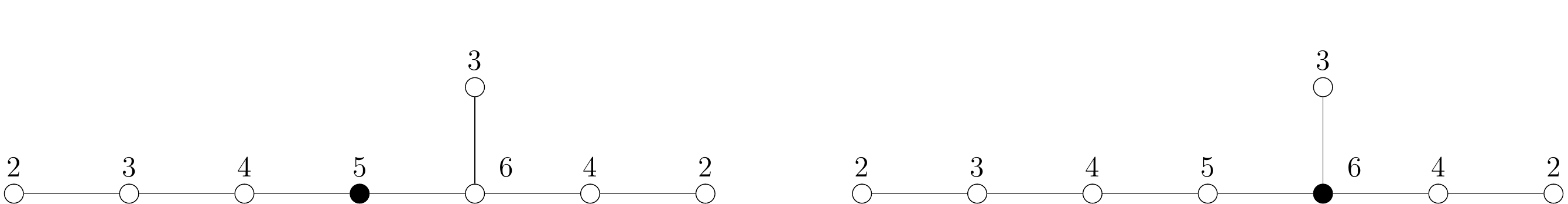}
\caption{The `colored' Dynkin diagrams $A_1,D_4,E_6,E_7,E_8$ corresponding to the chosen subgroup of the full Weyl group. (We consider two different colorings for $E_8$.) We will be quotienting the Cartan torus $\T$ by the Weyl group of the \emph{white} sub-diagram only.}
\label{fig:dynkins}
\end{figure}%
In the work of \cite{katz-morrison,curto-morrison}, one is studying partial resolutions of surfaces which involve a single $\pp^1$ only, so one chooses a particular node, such that only its corresponding sphere admits a simultaneous resolution, and keeps the rest shrunk at the origin. In other words, one has a single node colored in black, and Weyl subgroup in question is that of the subgraph in white in figure \ref{fig:dynkins}.\footnote{The same idea was subsequently exploited in \cite[Sec. 6]{cachazo-katz-vafa} to construct four-dimensional $\mathcal{N}=1$ gauge theories with adjoint matter by having a bunch of D5-branes wrap the blown-up sphere (the one corresponding to the black node), while the others are shrunk to zero size.}

In order to construct the special family admitting the partial simultaneous resolution, one takes the standard versal deformation, and writes the various coefficients as functions of the Cartan torus, such that these are invariant under the Weyl group of the (white) complementary Dynkin subgraph.


\section{Families of ALE surfaces from quivers} 
\label{sec:quivers}

In the previous section, we schematically explained that, in order to construct a family of ALE spaces admitting a simultaneous resolution, one writes the versally deformed ALE, but writes the deformation parameters as functions of the Cartan torus, partially quotiented under a suitably chosen subgroup of the full Weyl group.
While this approach is completely correct (and is the primary approach of \cite{katz-morrison}), it appears to be impractical for Lie algebras beyond $D_4$. 

Instead of directly generalizing \eqref{eq:A1def} at the level of hypersurface equations, we will use the approach put forth in \cite{karmazyn}. The idea is to reconstruct the geometry from a quiver with relations. The family of deformed ALE surfaces admitting a simultaneous resolution is indeed recovered as the relation satisfied by the gauge invariants of the former. This quiver is directly obtained starting from the Kronheimer quiver, i.e., the affine quiver of the ALE singularity. Below we summarize the results of \cite{karmazyn} that we will need in the following, and refer the reader to that paper for the detailed derivation. 

\begin{figure}[ht!]
\centering
\includegraphics[scale=1.25]{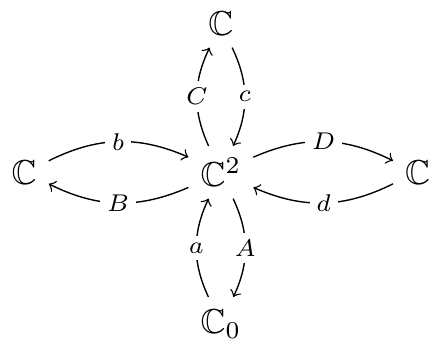}
\caption{The Dynkin--McKay quiver with relations reproducing the $D_4$ surface singularity $x^2=yz(y+z)$. The vertex labeled by a subscript $0$ corresponds to the affine node in the $\hat{D}_4$ Dynkin.}
\label{fig:D4Crep}
\end{figure}

We will exemplify the construction with the simplest nontrivial case, i.e.\ the $D_4$ singularity. Consider the (affine) $D_4$ McKay quiver in figure \ref{fig:D4Crep}. The figure actually depicts a \emph{representation of the quiver}, whereby each node corresponds to a vector space, and the arrows to linear maps among them. Notice that the dimension of each vector space is specified by the label in the colored $D_4$ Dynkin diagram of figure \ref{fig:dynkins} (with the affine node always corresponding to a one-dimensional space). We collect these dimensions in a  \emph{dimension vector} $\vec{d} = (d_0, \ldots, d_4) = (1,1,1,1,2)$. The arrows correspond to linear maps, i.e.\ matrices, which are required to satisfy the following relations (akin to the F-terms one would obtain by taking derivatives of a superpotential):
\begin{equation}\label{eq:D4preproj}
aA=bB=cC=dD=0\ , \quad Aa+Bb+Cc+Dd=0\ .
\end{equation}
We are using conventions such that the composition of maps runs from left to right; to accomplish that we take transposes, so that e.g. $a: \cc \to \cc^2$ is a row two-vector, $A: \cc^2 \to \cc$ a column two-vector, and so on. Thus by e.g. $Aa$ we mean the $2 \times 2$ matrix obtained via Kronecker product.

To describe the versal deformation at the quiver level, we first need to modify the above relations as follows:
\begin{subequations}\label{eq:D4preprojdef2}
\begin{gather}
aA=t_0 \ , \quad bB=t_1 \ , \quad cC=t_2 \ , \quad dD=t_3  \ , \\ Aa+Bb+Cc+Dd=-t_4 1_2\ .
\end{gather}
\end{subequations}
The $t_i$'s in \eqref{eq:D4preprojdef2} are complex numbers,\footnote{They can be thought of as complex Fayet--Iliopoulos parameters in $4d$ $\Ntwo$ language.} and correspond to coordinates on the Cartan torus $\T_{\hat{D}_4}$ of $\hat{D}_4$. Since we are using the affine $\hat{D}_4$ Dynkin diagram, rather than the non-affine one (as previously done for $A_1$), the $t_i$'s must satisfy the additional constraint\footnote{More generally, for any affine ADE quiver we have
\begin{equation}\label{eq:cartanbasis}
\sum_i \omega_i t_i = 0\ , \nonumber
\end{equation}
where the $\omega_i$ are the labels on the nodes in figure \ref{fig:dynkins}, and for the affine node we always put $\omega_0=1$.}
\begin{equation}\label{eq:cartanD4}
t_0 + \ldots + t_3 +2t_4 =0\ .
\end{equation}
The quiver representation is supplemented by \emph{stability parameters}, which can be thought of as real Fayet--Iliopoulos (FI) constants $\vec{\xi} = (\xi_0, \ldots ,\xi_4)$, subject to the relation
\begin{equation}
\vec{d} \cdot \vec{\xi} = 0\ .
\end{equation}
The space of possible representations subject to certain \emph{stability} criteria forms a continuous moduli space. These criteria essentially ensure that `bad' points are excluded (much like one excludes the origin of a vector space when making a projective space). See our companion paper \cite{collinucci-fazzi-valandro} for more detailed explanations.
The moduli space turns out to be the ALE surface itself. (This is essentially Kronheimer's construction \cite{kronheimer-ALE}.)  The $\xi_i$ correspond to the K\"ahler modulus of the $i$-th exceptional $\pp^1$ of a resolution, and the $t_i$ corresponds to the volume of the $i$-th deformed (i.e.\ non-holomorphic) sphere w.r.t. the complex structure $(2,0)$-form $\Omega$.

If we now define the following three gauge-invariant loops (which are just complex numbers),
\begin{equation}
x:=aBbCcA\ , \quad y:=aCcA\ , \quad z:=aBbA\ ,
\end{equation}
it is straightforward to show (by repeatedly applying \eqref{eq:D4preprojdef2}) that these satisfy the following hypersurface relation:
\begin{multline}\label{eq:D4fulldef}
x^2 + \frac{1}{4} t_0 (4 y+4 z -t_3^2+t_2^2+t_1^2+t_0^2)x = \\ yz(y+z) +  \frac{1}{4} yz(-t_3^2+t_2^2+t_1^2+t_0^2) -\frac{1}{16} t_0^2 t_1^2 t_2^2+\frac{1}{4} t_1^2 y^2+\frac{1}{4} t_2^2 z^2\ .
\end{multline}
This is a deformation of the $D_4$ singularity
\begin{equation}\label{eq:D4}
x^2=yz(y+z)\,.
\end{equation}
After a suitable coordinate redefinition, it can be brought to the well-known versal form
\begin{equation}\label{eq:D4fulldefweyl}
\tilde{x}^2+\tilde{y}^2\tilde{z}-\tilde{z}^3+\alpha_2(t_i) \tilde{z}^2+\alpha_1(t_i) \tilde{z} + \alpha_0(t_i) + \beta(t_i) y=0\ ,
\end{equation}
where the $\alpha_i$ and $\beta$ are Weyl-invariant functions of the Cartan torus coordinates $t_i$.

We have now recovered a versal deformation of the $D_4$ surface that admits a full simultaneous resolution. In terms of figure \ref{fig:dynkins}, we have colored all nodes black, so that the complementary Weyl group is trivial. Although the functions that appear in the hypersurface are Weyl-invariant, they are written in terms of coordinates that are fully Weyl-covariant. 

Now, we will study the situation with only the central node colored in black. This is the essence of the work in \cite{karmazyn}. The Weyl group of the white $D_4$ subdiagram in figure \ref{fig:dynkins} is $S_2 \times S_2 \times S_2$ (that is, the product of three copies of the Weyl group of $A_1$), and the correct base change is given by:
\begin{equation}\label{eq:baseD4}
(t_0,\ldots,t_3) \mapsto (T_0,\ldots,T_3)  = \left(\frac{t_0}{2},\frac{t_1^2}{4},\frac{t_2^2}{4},\frac{t_3^2}{4}\right)\ ,
\end{equation}
where we have eliminated $t_4$ in favor of $(t_0,\ldots,t_3)$ upon using \eqref{eq:cartanD4}. 

\begin{figure}[ht!]
\centering
\includegraphics[scale=1.25]{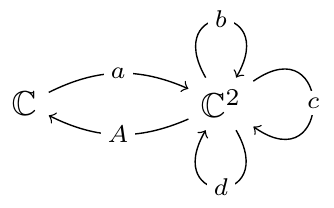}
\caption{The quiver with relations reproducing the threefold family of deformed $D_4$ surface singularities admitting a simultaneous resolution.}
\label{fig:length2}
\end{figure}

The procedure explained in \cite{karmazyn} entails considering a \emph{new} quiver obtained by `contracting' the external paths in figure \ref{fig:D4Crep} onto the central vertex, as depicted in figure \ref{fig:length2}. Notice that the new loops $b,c,d$ are represented by $2 \times 2$ matrices (rather than row vectors, as the notation in use would suggest).\footnote{The loops in figure \ref{fig:length2} \emph{are not} given by the same-name paths of figure \ref{fig:D4Crep}.} We also have the following new relations, inherited from \eqref{eq:D4preprojdef2}:
\begin{subequations}\label{eq:rellength2}
\begin{gather}
aA=2T_0\ , \quad b^2=T_1 1_2\ , \quad c^2=T_2 1_2\ , \quad d^2=T_3 1_2\ , \\ Aa+b+c+d=T_0 1_2\ .
\end{gather}
\end{subequations}
It is straightforward to show (upon repeatedly applying relations \eqref{eq:rellength2}) that the gauge invariants 
\begin{equation}\label{eq:D4gaugeinv}
x:=abcA\ , \quad y:=acA\ , \quad z:=abA
\end{equation}
satisfy the following relation:
\begin{multline}\label{eq:D4verdef}
x^2 + 2T_0(y+z -T_3+T_2+T_1+T_0^2)x = \\ yz(y+z) +  yz(-T_3+T_2+T_1+T_0^2) -4T_0^2T_1 T_2+T_1 y^2+ T_2 z^2\ ,
\end{multline}
which is again a deformation of the $D_4$ singularity \eqref{eq:D4}. So how is this different from the hypersurface in \eqref{eq:D4fulldef}? Both are fully deforming the singularity in the sense that, at a generic point on the Cartan torus, the ALE fiber is fully desingularized. However, whereas the hypersurface \eqref{eq:D4fulldef} admits a full simultaneous resolution, the one in \eqref{eq:D4verdef} only admits the simultaneous partial resolution that blows up the central node of the $D_4$ Dynkin diagram.

We have produced a non-compact CY \emph{sixfold}, if we take the whole family with the $T_i$'s promoted to coordinates. This family is known to be the so-called \emph{universal flop of length two} first constructed explicitly in \cite{curto-morrison} (albeit using a different method). Now we can build CY threefolds that admit a small resolution by simply taking a three-dimensional subspace of this sixfold (or an appropriate cover thereof).

For all other ALE cases, the prescription is exactly the same. The quivers producing the versal deformations of a Du Val surface singularity can be found in \cite[Sec. 4]{karmazyn}. They will make their appearance in section \ref{sec:gen-K3}.

\section{\texorpdfstring{$D_4$ singularity in ALE and universal flop of length two}{D4 singularity in ALE and universal flop of length two}} 
\label{sec:D4-quiv}

\begin{figure}[ht!]
\centering
\includegraphics[scale=.4]{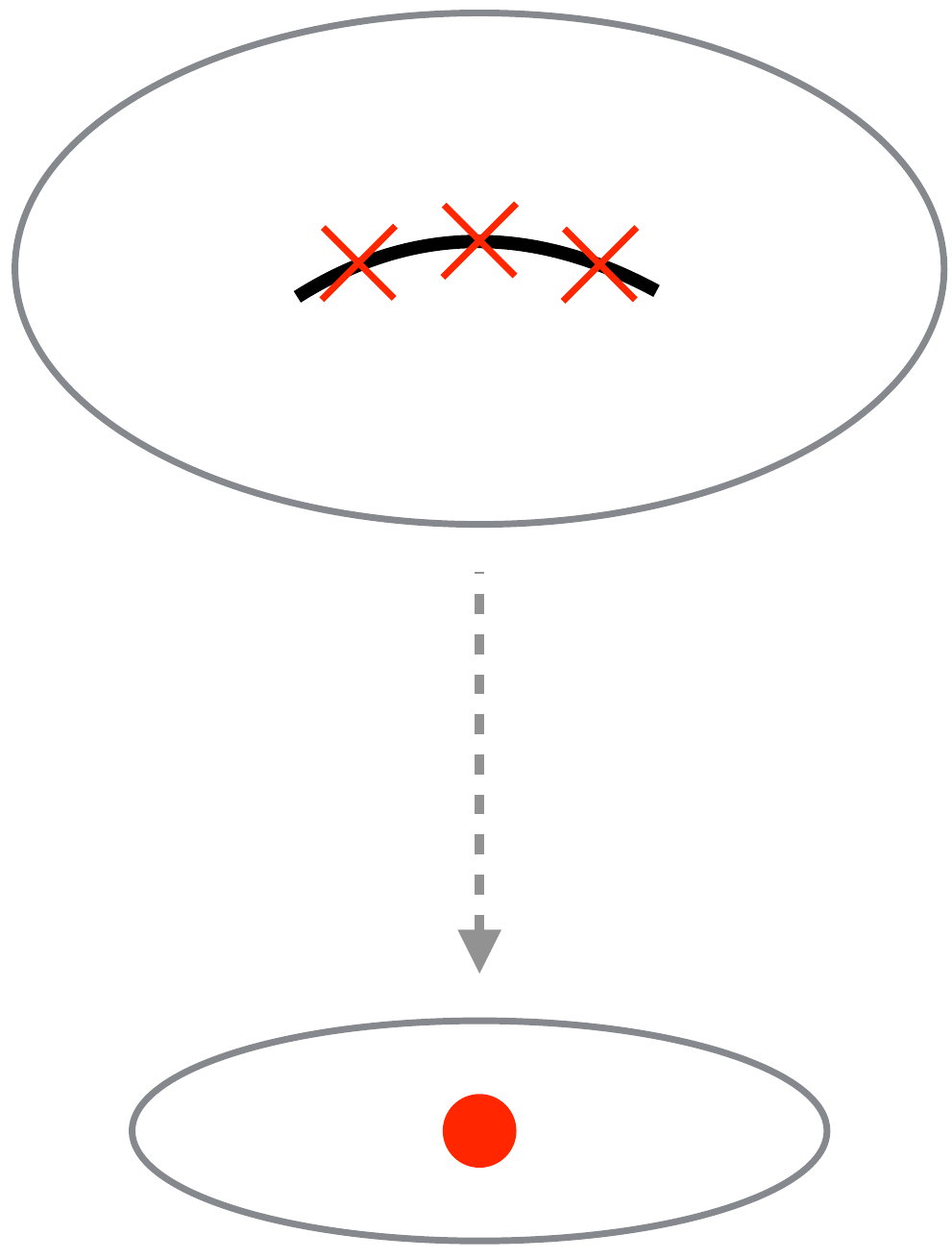}
\caption{Partial resolution of the sixfold family of versally deformed ALE spaces of type $\cc^2/D_4 =\Spec R$, with $R=\cc[x,y,z]/(x^2-yz(y+z))$. The base of the fibration is the deformation space parametrized by $(T_0,\ldots,T_3)$, whereas the fiber is parametrized by $(x,y,z)$. Over the red locus (which may be a nontrivial subvariety of the base) the fiber degenerates: We only resolve the black node in the colored $D_4$ Dynkin (see figure \ref{fig:dynkins}), corresponding to the two-dimensional vertex in figure \ref{fig:length2}. The resolution produces a single $\pp^1$ (the black line) with three singular $A_1$ points (the red crosses).}
\label{fig:singK3}
\end{figure}

In the previous section we have constructed a family of $D_4$ singularities, that is an ALE fibration over a four-dimensional base space, namely a sixfold. The ALE fiber is smooth over generic points of the base, and degenerates over codimension-one loci. In this section we want to describe this family from a different perspective, in order to understand why this model should generate charge-two matter in an M-theory compactification on a threefold slice (or appropriate cover of a slice) of the sixfold. We will first describe the ALE fibration over a cover of the base, using the Weyl-covariant coordinates. At the end we will quotient them by the proper Weyl subgroup, in order to have a fibration of the reduced $D_4$ quiver. See figure \ref{fig:singK3}.

\begin{figure}[ht!]
\centering
\includegraphics[scale=.4775]{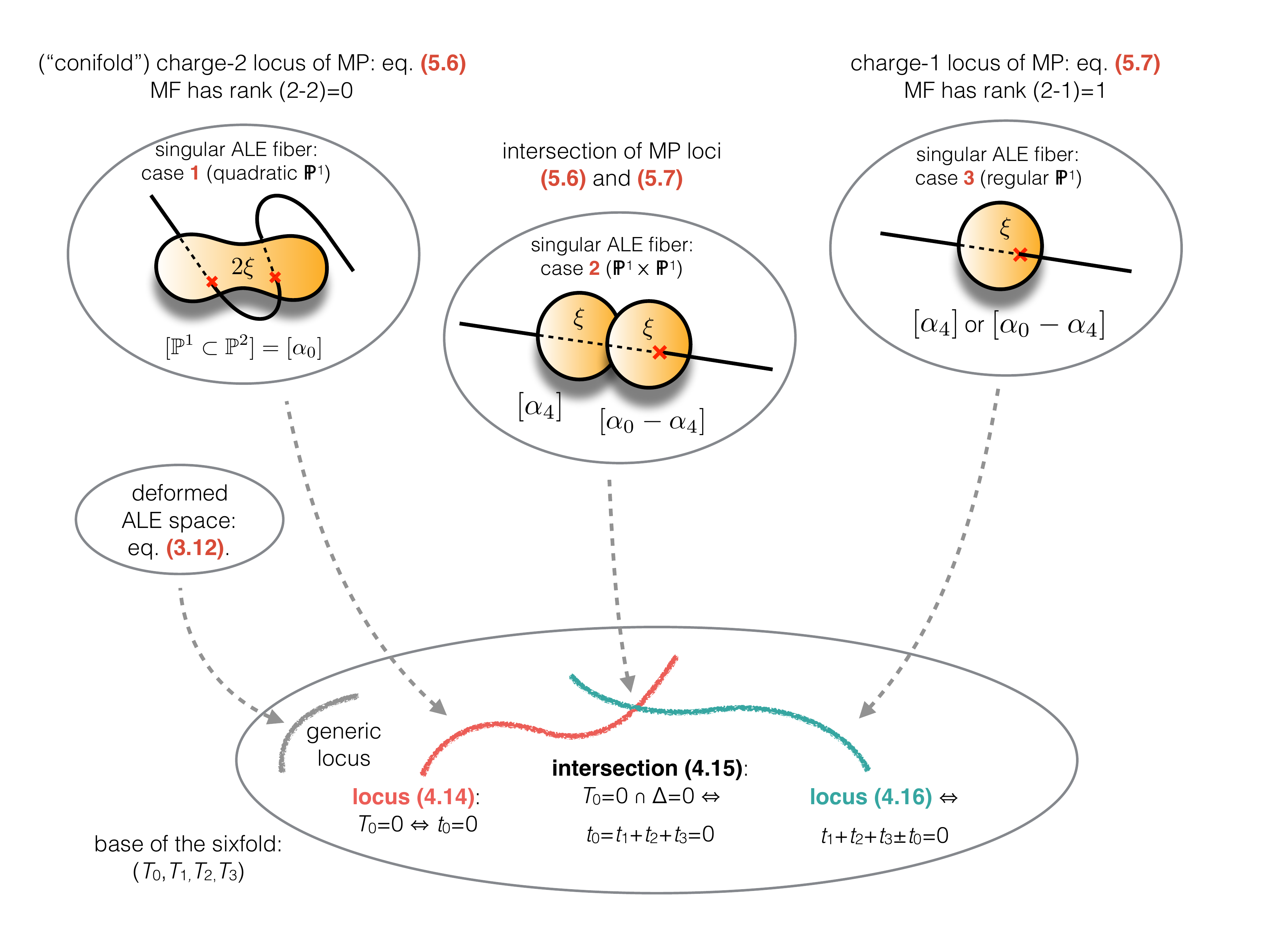}
\caption{Possible matter loci for charge-two examples. The cases referred to in the figure are \ref{locus1res}, \ref{locus2res}, and \ref{locus3res}. $(2)\xi$ is the real volume of the exceptional $\pp^1$ locus, $[\alpha_i]$ its homology class. The sixfold family is given by \eqref{eq:D4verdef}.}
\label{fig:P1s}
\end{figure}

Let us summarize this section. First, in section \ref{subsec:K3homology}, we briefly introduce the hyper-K\"ahler structure of ALE spaces in a notation that will serve us. We explain in terms of the complex structure form and the K\"ahler form what it means to have a `vanishing sphere'.
In section \ref{subsec:DefK3}, we discuss the deformations of ALE spaces. We single out three special subspaces in the parameter space of deformations that will be of interest: Two of them will display residual $A_1$ singularities; the third one is given by the intersection of the first two and it will display an $A_2$ singularity.
In section \ref{subsec:ResK3}, we show how these residual singularities are resolved. Here, we will realize that the two $A_1$ loci  show us the various types of charged matter that can appear. For one of the $A_1$ singularities, the exceptional $\pp^1$ is intersected once by the $\U(1)$ divisor and gives rise to a charge-one hypermultiplet. 
The other type of $A_1$ singularity has a different structure, it corresponds to a $\pp^1$ that is quadratically embedded into the geometry, and therefore twice intersected by the $\U(1)$ divisor. This  gives rise to a charge-two hyper.
See figure \ref{fig:P1s}.

\subsection{\texorpdfstring{$D_4$ singularity on an ALE space}{D4 singularity on an ALE space}}
\label{subsec:K3homology}

An ALE space develops an ADE singularity when a bunch of two-spheres in its homology collapse to zero size. If the intersection form of the surface restricted to these spheres is (minus) the Cartan matrix of an ADE algebra $\Gamma_\text{ADE}$, then the surface has a singularity of type $\Gamma_\text{ADE}$.

The volume of the homologically non-trivial cycles is measured by the metric. The metric of any ALE space is determined by three harmonic two-forms, that span a three-plane $\Sigma$ inside  $H^{2}(\text{surface})$. The latter is a vector space equipped with a natural scalar product defined as
\begin{equation}\label{K3modmetric}
v\cdot w := \int_\text{surface} v\wedge w \ , \quad\quad v,w\in H^2(\text{surface}) \ .
\end{equation}
This product must be positive definite when restricted to $\Sigma$. In terms of Poincar\'e dual two-cycles, the product is simply given by the intersection number.

The ALE surface is a local Calabi--Yau (i.e., it is Ricci flat), whose metric is determined by the K\"ahler form and the holomorphic $(2,0)$-form $\Omega$, which satisfy
\begin{align}\label{OmegaJconstr}
\Omega \cdot \Omega = 0\ , \quad J \cdot \Omega = 0\ , \quad
\Omega \cdot \bar{\Omega}>0\ , \quad J \cdot J >0\ .
\end{align}
These two forms can be constructed by choosing three orthogonal vectors $\omega_1$, $\omega_2$ and $\omega_3$ spanning $\Sigma$: 
\begin{equation}
\Omega = \omega_1 + i \omega_2 \qquad\mbox{ and }  \qquad  J=\omega_3\:.
\end{equation}
The metric is invariant under $\SO(3)$ rotations of the $\omega_i$'s. By such a rotation, one can change the choice of $\Omega$ and then of a complex structure. In fact, any hyper-K\"ahler manifold has a whole $S^2$ worth of complex structures. 

The position of $\Sigma$ in $H^2(\text{surface}, \zz)$ determines what cycles have zero size: If a two-cycle is orthogonal to $\Sigma$, its volume is zero.\footnote{By this we actually mean $\int_{\alpha}\Omega = \int_{\alpha}J=0 $. 
Since two-forms are Poincar\'e dual to two-cycles, we often write these conditions as an orthogonality condition with $\Sigma$, i.e.\ $\alpha\cdot \Omega=\alpha\cdot J=0$.}
 By adjunction, one can verify that the classes of two-spheres have self-intersection $-2$. Hence an ADE singularity of type $\Gamma_\text{ADE}$ is present when $\Sigma$ is orthogonal to a set of two-cycles with intersection matrix equal to minus the Cartan matrix of the ADE group $\Gamma_\text{ADE}$ (in particular this means that they have self-intersection $-2$).

In this section, we are interested in a hyper-K\"ahler surface with a $D_4$ singularity. Following the general rule, this happens when $\Sigma$ is orthogonal to four independent spheres $\alpha_1$, $\alpha_2$, $\alpha_3$, $\alpha_4$  in $H^2(\text{surface},\mathbb{Z})$ that have the following intersection pattern:
\begin{equation}
 \alpha_i\cdot\alpha_j = C_{ij} \qquad\text{with}\qquad C = \begin{pmatrix}
 -2 & & & 1 \\   & -2 & & 1 \\    & & -2 & 1 \\   1 & 1 & 1 & -2  \end{pmatrix}\ .
\end{equation}
We will call 
\begin{equation}
V^{D_4} = \langle \alpha_1, \alpha_2, \alpha_3, \alpha_4 \rangle
\end{equation}
the four-dimensional subspace spanned by the spheres $\alpha_i$, that we will also call simple roots. All vectors in $V^{D_4}$ that square to $-2$ represent homology classes of two-spheres inside the surface corresponding to the roots of $D_4$.\footnote{
Let us list all twelve positive roots of $D_4$:
$
\alpha_1,\alpha_2,\alpha_3,\alpha_4,\alpha_1+\alpha_4,\alpha_2+\alpha_4,\alpha_3+\alpha_4,\alpha_1+\alpha_2+\alpha_4,\alpha_1+\alpha_3+\alpha_4,\alpha_2+\alpha_3+\alpha_4,\alpha_1+\alpha_2+\alpha_3+\alpha_4,\alpha_1+\alpha_2+\alpha_3+2\alpha_4
$.
}
The simple roots $\alpha_i$ (and consequently all the roots) have zero size when they are orthogonal to $\Sigma$, i.e.\ when
\begin{equation}
 \alpha_i\cdot\Omega^{D_4} = \alpha_i\cdot J^{D_4}=0\ , \qquad \forall i=1,\ldots,4\ .
\end{equation}

\subsection{\texorpdfstring{Deformation of the $D_4$ singularity}{Deformation of the D4 singularity}}
\label{subsec:DefK3}


To smooth out the $D_4$ singularity one needs to move the plane $\Sigma$ inside $H^2(\text{surface})$ such that $\Omega$ or $J$ are no longer orthogonal to the $\alpha_i$. We say that we {\it deform} the singularity if we keep $J$ orthogonal to $V^{D_4}$, while letting  $\Omega$ have components along $V^{D_4}$:
\begin{equation}
 J=J^{D_4} \ , \qquad \Omega = \Omega'^{D_4} + t_1\alpha_1^\ast + t_2\alpha_2^\ast + t_3\alpha_3^\ast + t_4\alpha_4^\ast  \ .
\end{equation}
Here $\Omega'^{D_4}$ is a $(2,0)$-form that is still orthogonal to $V^{D_4}$ and is chosen such that \eqref{OmegaJconstr} are  satisfied by $J$ and the new $\Omega$.\footnote{Take $\omega_m^0$ such that $\omega_m^0\cdot\omega_k^0=\delta_{mk}$ and $\omega_m\cdot\alpha_i=0$ $\forall m,i$. Then one can write $J^{D_4}=\nu \omega_3^0$ and $\Omega^{D_4}=\omega_1^0+i\omega_2^0$. One can take $\Omega'^{D_4}=\Omega^{D_4}+b\omega_1^0$, with $b=-\sum_{ij}t_it_j(\alpha_i^\ast\cdot\alpha_j^\ast)$.} 
 $t_i$ ($i=1,...,4$) are complex numbers. $\{\alpha_i^\ast\}$ is a dual basis of $\{\alpha_i\}$ in $V^{D_4}$, i.e.\ they are such that $\alpha_i^\ast\cdot\alpha_j=\delta_{ij}$.
For convenience, we write down the expression of $\alpha_i^\ast$ in terms of $\alpha_j$:
\begin{subequations}
\begin{align}
\alpha_1^\ast &= -\alpha_1-\alpha_4-\tfrac12 (\alpha_2+\alpha_3) \ ,\\
\alpha_2^\ast &= -\tfrac12( \alpha_1+ 2\alpha_2+ \alpha_3+ 2\alpha_4) \ ,\\
\alpha_3^\ast &= -\tfrac12( \alpha_1+ \alpha_2+ 2\alpha_3+ 2\alpha_4) \ ,\\
\alpha_4^\ast &= -\alpha_1- \alpha_2- \alpha_3- 2\alpha_4 \ .
\end{align}
\end{subequations}
The ALE fibration over the space with coordinates $t_i$ is the sixfold family that we want to construct up to an appropriate quotient by a subgroup of the Weyl group of $D_4$.

With a generic choice of $t_i$'s, one obtains a non-zero volume for all the (integral) spheres in $V^{D_4}\subset H_2(\text{surface})$ and the resulting ALE space is smooth. 
There are however choices of these parameters that leave some sphere in $V^{D_4}$ orthogonal to $\Sigma$. This means that the ALE space is still singular. Let us consider the following relevant examples (where we use relation \eqref{eq:cartanD4}):
\begin{enumerate}
\item \label{locus1} $\mathbf{-t_0 = t_1+t_2+t_3+2t_4=0}$. The vector 
\begin{equation}
\alpha_0 := \alpha_1+\alpha_2+\alpha_3+2\alpha_4
\end{equation}
has self-intersection $-2$ and is orthogonal to $\Omega$. If the values of the $t_i$'s are the most generic ones satisfying this condition, all the other spheres in $V^{D_4}$ have finite size (i.e.\ are not orthogonal to $\Omega$). We then have an $A_1$ singularity.
\item \label{locus2} $\mathbf{-t_0 = t_1+t_2+t_3+2t_4=0,\, \pm t_1\pm t_2\pm t_3=0}$. We have three  vectors in $V^{D_4}$ that have zero size. In the case $+++$ they are
\begin{equation}
\alpha_4\ , \quad  \alpha_0-\alpha_4\ , \quad \alpha_0\ .
\end{equation}
They span a two-dimensional subspace of $V^{D_4}$: take $\alpha_4$ and $\alpha_0-\alpha_4$ as a basis, they intersect at one point and give a choice of simple roots for $A_2$. The ALE space develops an $A_2$ singularity. Different choices of sign in $\pm t_1\pm t_2\pm t_3$ select a different set of three roots that shrink; they always have the right intersection pattern to give an $A_2$ singularity.
\item \label{locus3} $\mathbf{ \pm t_1\pm t_2\pm t_3\pm t_0=0,\,t_0\neq 0}$. There is only one sphere that has zero volume. In the case $++++$ it is
\begin{equation}
\alpha_4\ .
\end{equation}
We have an $A_1$ singularity. For the choice $+++-$ the zero-size sphere is $\alpha_0-\alpha_4$.
\item \label{locus4} $\mathbf{t_1=t_2=t_3=t_4=0}$. All the spheres in $V^{D_4}$ have zero size. We have the $D_4$ singularity.
\end{enumerate}
Notice that the second case is the intersection of the first and third.

\subsection{Resolutions of remaining singularities}
\label{subsec:ResK3}

Over some loci of the $t_i$ space, the ALE space is still singular. Over these loci one can resolve the singularities by moving $J$  in such a way that the spheres orthogonal to $\Omega$ are not orthogonal to $J$ anymore. This means that $J$ now takes the form
\begin{equation}
J=J'^{D_4} + \sum_{i=1}^4\xi_i\alpha_i^\ast\ ,
\end{equation}
where $J'^{D_4}$ is a two-form that is still orthogonal to $V^{D_4}$ and is chosen such that \eqref{OmegaJconstr} are still satisfied. The $\xi_i$'s are real numbers.

The procedure for constructing partial simultaneous resolutions has led us to replace the original $D_4$ quiver by a contracted quiver with only two nodes. See figure \ref{fig:length2}. In quiver representation language, this means that we really only have two stability parameters (i.e. K\"ahler moduli) at our disposal, namely $\xi_0$ and $\xi_4$. Moreover, these must satisfy $2\xi_0+\xi_4=0$. In the language of D3-probes at singularities, we only have one overall $\U(1)$ in the quiver, so only one real FI parameter.

Therefore, we are interested in resolutions consistent with the contracted $D_4$ quiver, where we allow simultaneous resolution only along the central root of the $D_4$ Dynkin diagram. This  corresponds to fiberwise resolutions of the following form for $J$:
\begin{equation}\label{JamputQuiv}
J=J'^{D_4} + \xi\,\alpha_4^\ast\ ,
\end{equation}
i.e.\ we allow only $\xi := \xi_4$ to be non-zero (while $\xi_{1,2,3}=0$).

Let us consider the previous examples and see which spheres acquire a non-zero volume when $J$ is given by \eqref{JamputQuiv}:
\begin{enumerate}
\item \label{locus1res} $\mathbf{-t_0:= t_1+t_2+t_3+2t_4=0}$. The root $\alpha_0$ gets volume 
\begin{equation}
J\cdot \alpha_0=2\xi\ .
\end{equation}
\item \label{locus2res} $\mathbf{-t_0:= t_1+t_2+t_3+2t_4=0,\, \pm t_1\pm t_2\pm t_3=0}$. Let us concentrate on the case $+++$:
The simple roots $\alpha_4$ and $\alpha_0-\alpha_4$ have volume 
\begin{equation}
J\cdot \alpha_4=J\cdot (\alpha_0-\alpha_4)=\xi\ .
\end{equation}
Notice that the third root is $\alpha_0=\alpha_4+(\alpha_0-\alpha_4)$ that has volume $2\xi$ and is homologous to the sphere discussed at point \ref{locus1res}.
\item \label{locus3res} $\mathbf{ \pm t_1\pm t_2\pm t_3\pm t_0=0,\,t_0\neq 0}$. Let us consider the case with $++++$: the root $\alpha_4$ gets a volume 
\begin{equation}
J\cdot \alpha_4=\xi\ .
\end{equation}
In the case $+++-$, we have $J\cdot (\alpha_0-\alpha_4)=\xi\:.$
\item \label{locus4res} $\mathbf{t_1=t_2=t_3=t_4=0}$. Only the spheres that are linear combinations of $\alpha_4$ with some other sphere get non-zero size.
\end{enumerate}
Consider the ALE fibration over the space with coordinates $t_1,t_2,t_3,t_4$. Over generic points of the base, the ALE fiber is smooth. Over the locus $t_0=0$  there is an $A_1$ singularity whose resolution gives an exceptional $\mathbb{P}^1$ in the class $\alpha_0$ (case \ref{locus1res}). On top of $t_1+t_2+t_3+t_0=0$ the resolved $\mathbb{P}^1$ is in the class $\alpha_4$, while on top of $t_1+t_2+t_3-t_0=0$ it is in the class $\alpha_0-\alpha_4$ (case \ref{locus3res}).
All these three loci intersect at $t_0=t_1+t_2+t_3=0$, where we have two $\mathbb{P}^1$'s intersecting at one point (case \ref{locus2}). Case \ref{locus4res} is special, and will be treated separately in sections~\ref{sub:mf} and \ref{sub:laufer}.

Notice that the fact that the class $[\alpha_0]$ has volume $2 \xi$ is compatible with the fact that this is a quadratically embedded $\pp^1$. We can think of it as follows: Take case \ref{locus2res}, which has a union of two $\pp^1$'s, one of class $\alpha_4$, and one of class $\alpha_0-\alpha_4$. If we deform this locus, the two spheres will coalesce into a single sphere of class $\alpha_0-\alpha_4+\alpha_4 = \alpha_0$. Hence, $\alpha_0$ is a sphere that is the homological sum of two projective lines. From this picture, we understand that it gets intersected twice by the Weil divisor corresponding to the small resolution, and hence gives matter of charge two.

All of the homology and volume relations we have discovered can be phrased
in terms of the global Picard group and the local Picard groups of 
the singularities.  For example, the fact that the volume of the root
$\alpha_0$ is twice the volume of the roots $\alpha_4$ and $\alpha_0-\alpha_4$
is the statement that the class of $\alpha_0$ is twice the generator
of the Picard group, while the classes of $\alpha_4$ and $\alpha_0-\alpha_4$
correspond to generators.


\subsection{Algebraic description} 
\label{sec:AlgebrDescr}

We now relate the above description of the ALE-fibered sixfold with the algebraic description given in section~\ref{sec:quivers}. The coordinates $T_i$ of the four-dimensional base are expressed in terms of the covering coordinates $t_i$ according to \eqref{eq:baseD4}.
Remember that $t_0 = -(t_1+t_2+t_3+2t_4)$.  
Let us repeat the equation describing the generic (deformed) ALE fiber:\footnote{This is nothing but \eqref{eq:D4verdef} with $x^\text{here} =x^\text{there} + \mathrm{v}$, where $\mathrm{v}$ (introduced below) is (half) the coefficient of the linear term in $x^\text{there}$.}
\begin{multline}\label{deformedD4amputquiv}
x^2=T_0^2(y+z-T_3+T_2+T_1+T_0^2)^2\ + \\ +(y+z-T_3+T_2+T_1+T_0^2)yz-4T_0^2T_1T_2+T_1y^2+T_2z^2\ ,
\end{multline}
where $x,y,z$ are coordinates along the ALE fiber. The sixfold \eqref{deformedD4amputquiv} is singular at the vanishing loci of the following two ideals:
\begin{equation}\label{SingLocCh2}
 (x,\, y,\,z,\,T_0) 
\end{equation}
and
\begin{equation}\label{SingLocCh1}
(x,\,\,\, \mathrm{v}\,y+2T_2z,\,\, \,2T_1y+\mathrm{v}\,z, \,\,\, y^2-4T_2T_0^2,\,\,\, z^2-4T_1T_0^2,\,\, \, zy+2T_0^2\mathrm{v}, \,\,\, 4T_1T_2-\mathrm{v}^2)\ ,
\end{equation}
with $\mathrm{v}:= (z+y+T_1+T_2-T_3+T_0^2)$.\footnote{We notice that $\mathrm{v}=-2v$, where $v$ is the variable appearing in the universal flop of length two of \cite{curto-morrison}, that we reproduce in \eqref{eq:univ}.}

The quiver in figure \ref{fig:length2} provides a (simultaneous) resolution of these singularities. As we are now going to show, over the two loci the exceptional locus is a $\mathbb{P}^1$. At the intersection of the two loci, the exceptional locus is the union of two intersecting $\mathbb{P}^1$'s. In the next section we will show that these $\mathbb{P}^1$'s correspond to the blown-up spheres of the ALE fiber  found above.

\subsubsection{Exceptional locus}\label{sec:ExcepP1}

We start by restricting ourselves over the locus $T_0=0$ (i.e.\ $t_0=0$) \eqref{SingLocCh2}, where we know the ALE fiber should have an $A_1$ singularity. In the resolution given by the quiver, we should see an exceptional $\mathbb{P}^1$ over this locus.

Over $T_0=0$ the relations \eqref{eq:rellength2} become:
\begin{equation}
aA=0\ , \quad b^2 = T_1 1_2\ , \quad c^2 = T_2 1_2\ , \quad d^2 = T_3 1_2\ , \quad Aa+b+c+d=0 \ .
\end{equation}
First of all notice that the stability conditions imply that, at $x=y=z=0$, one has $A=0$.\footnote{
Remember that $a, ab, ac, abc$ must generate the whole $\mathbb{C}^2$ vector space at the right vertex (this is equivalent to insuring that `destabilizing' quiver representations are excluded \cite{collinucci-fazzi-valandro}). Hence, the condition $x=y=z=0$, with $x,y,z$ given in \eqref{eq:D4gaugeinv}, force $A=0$: in fact, $A$ is implied to be orthogonal to a complete set of vectors of $\mathbb{C}^2$.} Moreover,
we can use the last relation to eliminate $d$. We are then left with:
\begin{subequations}
\begin{align}
\label{rel:beta2}b^2 &= T_1 1_2\ ,\\
\label{rel:gamma2}c^2 &= T_2 1_2\ ,\\
\label{rel:betagamma}\{b,c\} &= (T_3 -T_1 -T_2)1_2\ .
\end{align}
\end{subequations}
The matrices $b,c$ can be expanded as
\begin{equation}
 b = \beta_0 1_2 + \beta_i \sigma^i\ , \quad c = \gamma_0 1_2 + \gamma_i \sigma^i\ .
\end{equation}
Let us consider the relation \eqref{rel:beta2}. It says that $b^2$ must be proportional to the identity matrix, i.e.\
\begin{equation}
b^2=(\beta_0 1_2 + \beta_i \sigma^i )^2 = (\beta_0^2+\beta_i^2)1_2 + 2\beta_0\beta_i \sigma^i  \propto 1_2\ .
\end{equation}
This happens if and only if  $\beta_0=0$ or $\beta_i=0$ $\forall i=1,2,3$. The second option is not possible: If the $\beta_i$'s vanish, $b\propto 1_2$ and this would also imply $c\propto 1_2$ (see \eqref{rel:betagamma}); however this is forbidden by the stability condition (i.e.\ $\text{Span}\langle a, ab , ac ,a b c \rangle\cong \cc^2$ must be two-dimensional). Therefore we must take $\beta_0=\gamma_0=0$, i.e.\ 
\begin{equation}
 b =  \beta_i \sigma^i \  , \quad c = \gamma_i \sigma^i\ .
\end{equation}
Plugging this into \eqref{rel:beta2}, \eqref{rel:gamma2}, and \eqref{rel:betagamma} one obtains:
\begin{equation}\label{relb2c2bc}
 \sum_i \beta_i\beta_i = T_1\ , \quad \sum_i \gamma_i\gamma_i=T_2\ , \quad \sum_i \beta_i\gamma_i = \frac12 (T_3 -T_1 -T_2)\ .
\end{equation}
It is now possible to construct three $\text{SL}(2,\mathbb{C})$-invariant object (where $\text{SL}(2,\mathbb{C})$ is acting on the right node of the quiver):
\begin{equation}
 s_1 := a\wedge ab \, \quad  s_2 := a\wedge ac \ , \quad  s_3 := a\wedge abc \ .
\end{equation}
Because of the stability condition, these cannot vanish simultaneously. Moreover, they are charged under the relative $\mathbb{C}^\ast$ between the two nodes. Hence they are coordinates on a $\mathbb{P}^2$. There is a homogeneous quadratic relation among the $s_i$, which selects a $\mathbb{P}^1$ inside $\mathbb{P}^2$. It is given by
\begin{equation}\label{quadraticrelsi}
 s_3^2 = T_2 s_1^2 - (T_3-T_1-T_2)s_1s_2 + T_1 s_2^2\ .
\end{equation}
\emph{Proof}. First, notice that 
\begin{equation}
 bc=\beta_i\gamma_j \sigma^i\sigma^j = \beta_i\gamma_j (\delta^{ij}1_2 + i \epsilon^{ijk}\sigma^k)\ .
\end{equation}
Hence
\begin{equation}
s_3 = \beta_i\gamma_j (\delta^{ij}a\wedge a + i \epsilon^{ijk}a\wedge a\sigma^k) = i\,\beta_i\gamma_j \epsilon^{ijk}a\wedge a\sigma^k
\end{equation}
For convenience, let us call $X^k := a\wedge a\sigma^k$. Then 
\begin{equation}
 s_1 = \beta_iX^i \ ,\quad  s_2 =\gamma_i X^i \ , \quad  s_3 = i \,\beta_i\gamma_j \epsilon^{ijk} X^k \ .
\end{equation}
Using the relation
\begin{equation}
 \epsilon^{ijk}\epsilon^{h\ell p} = \delta^{ih}(\delta^{j\ell}\delta^{kp}-\delta^{jp}\delta^{k\ell}) -  \delta^{i\ell}(\delta^{jh}\delta^{kp}-\delta^{jp}\delta^{kh}) +  \delta^{ip}(\delta^{jh}\delta^{k\ell}-\delta^{j\ell}\delta^{kh})\:,
\end{equation}
we can finally write
\begin{align}
s_3^2 &= - \beta_i\gamma_j\beta_h\gamma_\ell \,\epsilon^{ijk}\epsilon^{h\ell p} \,X^kX^p \nonumber\\
 &= \left( ((\vec{\beta}\cdot\vec{\gamma})^2  - {\vec{\beta}}^{\,2}{\vec{\gamma}}^{\,2} )\delta^{kp}  +\vec{\gamma}^2\beta_k\beta_p -   (\vec{\beta}\cdot\vec{\gamma}) (\beta_k\gamma_p+\beta_p\gamma_k) + \vec{\beta}^2 \gamma_k\gamma_p \right)X^kX^p
\nonumber \\
 &= \vec{\gamma}^2 (\beta_k X^k)^2 - 2(\vec{\beta}\cdot\vec{\gamma}) (\beta_kX^k)(\gamma_pX^p) + \vec{\beta}^2 (\gamma_pX^p)^2 \\
 &=\vec{\gamma}^2 s_1^2 - 2 (\vec{\beta}\cdot\vec{\gamma}) s_1s_2 + \vec{\beta}^2 s_2^2 \nonumber\\
 &= T_2 s_1^2 -  (T_3-T_1-T_2) s_1s_2 + T_1 s_2^2 \nonumber\ ,
\end{align}
where we used $\sum_k X^kX^k=0$ and the relations \eqref{relb2c2bc}. $\square$ 

We have thus shown that the exceptional $\mathbb{P}^1$ over the locus $T_0=0$ is given by the quadratic equation \eqref{quadraticrelsi} inside $\mathbb{P}^2[s_1:s_2:s_3]$. 

We are now interested in following the fate of this $\mathbb{P}^1$ when the locus $T_0=0$ intersects the second singular locus in the sixfold, namely \eqref{SingLocCh1}. This happens when 
\begin{equation}\label{Deltatbtctd}
\Delta :=(T_3-T_1-T_2)^2-4T_1T_2 = 0\ .
\end{equation}
Looking at equation \eqref{quadraticrelsi}, one sees that at this locus the exceptional $\mathbb{P}^1$ splits into two $\mathbb{P}^1$'s. By a gauge fixing, we now show that the exceptional locus at $T_0=0$ is exactly parametrized by a $\mathbb{P}^1$. We started from coordinates $\beta_i,\gamma_i,a_1,a_2$ modulo the relations \eqref{relb2c2bc} and modulo (complexified) gauge transformations $(\mathbb{C}^\ast \times \text{GL}(2,\mathbb{C}))/\mathbb{C}^\ast$.
We first break $\text{SL}(2,\mathbb{C})$ to the $\mathbb{C}^\ast$ generated by the Cartan, by choosing
\begin{equation}
 b= T_1^{1/2} \sigma_3 \ , \quad c=\begin{pmatrix} r & q  \\  p & -r  \end{pmatrix}\ ,
\end{equation}
and putting $\gamma_1=\tfrac12(q+p)$, $\gamma_2=\tfrac{i}{2}(q-p)$, $\gamma_3=r$. The relations \eqref{rel:betagamma}  and \eqref{rel:gamma2} now impose
\begin{equation}\label{relpqr}
p\,q= -\frac{\Delta}{4T_1}   \ ,  \quad  r^2  = \frac{(T_3-T_1-T_2)^2}{4T_1}\ ,
\end{equation}
with $\Delta$ as in \eqref{Deltatbtctd}.

When $\Delta\neq 0$, we can use the $\mathbb{C}^\ast$ generated by the Cartan to fix $q=p$, leaving behind only the relative $\mathbb{C}^\ast$ (between the two nodes of the quiver); the relations  \eqref{relpqr} then determine $c$ completely. The exceptional locus is parametrized by the components of the row two-vector $a$, i.e.\ $(a_1,a_2)$, modulo the relative $\mathbb{C}^\ast$ action. These actually span a $\mathbb{P}^1[a_1:a_2]$, since the stability condition excludes the point $(a_1,a_2)=(0,0)$.

When $\Delta=0$, the first equation in \eqref{relpqr} factorizes, giving two loci, one at $p=0$ and one at $q=0$. Let us consider the former: the $\mathbb{C}^\ast$ generated by the Cartan can be fixed by choosing $q=T_3$. Again, we are left with a $\mathbb{P}^1$ parametrized by $[a_1:a_2]$. We can do the same with the locus $q=0$, obtaining a second $\mathbb{P}^1$. We have then explicitly shown that at $\Delta=0$ the exceptional $\mathbb{P}^1$ splits into two $\mathbb{P}^1$'s, intersecting at $p=q=0$. (At this point we still need to fix the Cartan $\mathbb{C}^\ast$; we do this by choosing $a_1\propto a_2$, i.e.\ we get a point.)

\subsection{Correspondence with sections~\ref{subsec:DefK3} and~\ref{subsec:ResK3}} 
\label{sec:karma}

We now show that the exceptional $\mathbb{P}^1$'s found in section~\ref{sec:ExcepP1} are exactly the spheres over the loci called \ref{locus1res} and \ref{locus3res} in sections~\ref{subsec:DefK3} and~\ref{subsec:ResK3}.

Let us begin with the locus \eqref{SingLocCh2}. In terms of the parameters $t_i$'s this means that $t_0 =  -( t_1+t_2+t_3+2t_4)=0$, i.e.\ it corresponds to locus \ref{locus1res}, where the resolved sphere was $\alpha_0$ with volume $2\xi$. In section~\ref{sec:ExcepP1} we found that the exceptional $\mathbb{P}^1$ is a quadric in $\mathbb{P}^2$ that splits when $(T_3-T_1-T_2)^2=4T_1T_2$. In terms of the $t_i$'s (using the relations \eqref{eq:baseD4}) the latter condition becomes
\begin{equation}
0=T_3-T_1-T_2\mp 2T_1^{1/2}T_2^{1/2}=\tfrac14(t_2^2-t_1^2-t_2^2\mp2t_1t_2)=\tfrac14 t_3^2- \tfrac14(t_1\pm t_2)^2\ .
\end{equation}
Hence we have 
\begin{equation}\label{t1t2t3pm}
 \pm t_1 \pm t_2 \pm t_3 =0\ .
\end{equation}
We recognize the case \ref{locus2res}, where the exceptional fiber was made up of two spheres (in the classes $\alpha_4$ and $\alpha_0-\alpha_4$ when we choose $+++$), both with volume $\xi$. The choice of $\pm$ sign is only available on the covering space parametrized by the $t_i$: Once we quotient by the appropriate Weyl subgroup (i.e.\ we pass to the Weyl-invariant coordinates $T_i$) the loci \eqref{t1t2t3pm} become branches of the same locus.

We now analyze the  singular locus \eqref{SingLocCh1}. We first see that the singularity occurs at non-zero $y$ and $z$:
\begin{equation}
y= \pm 2T_2^{1/2}T_0 = \pm \frac12 t_2 t_0 \ , \quad z= \pm 2T_1^{1/2}T_0 = \pm \frac12 t_1 t_0 \ .
\end{equation}
The other independent relation in the ideal \eqref{SingLocCh1} is $\mathrm{v}^2-4T_1T_2=0$, that in terms of the $t_i$'s reads
\begin{align}
0 &= \mathrm{v} \pm 2T_1^{1/2}T_2^{1/2} \nonumber\\
  &= (\pm 2T_1^{1/2}T_0 \pm 2T_2^{1/2}T_0 + T_1+T_2-T_3+T_0^2) \pm 2T_1^{1/2}T_2^{1/2} \nonumber \\
  &= \frac14(\pm 2t_1t_0\pm 2t_2t_0 + t_1^2+t_2^2-t_3^2+t_0^2 \pm 2t_1t_2 ) \\
  &=\frac14(t_1\pm t_2 \pm t_0)^2 -\frac14 t_3^2 \nonumber\ .
\end{align}
Hence, we have
\begin{equation}
 \pm t_1 \pm t_2 \pm t_3 \pm t_0 = 0\ ,
\end{equation}
with $t_0$ generically different from zero. We obtain all the branches of the locus \ref{locus3res} in the base: Choosing $+++\pm$, the exceptional $\pp^1$ is either in the class $\alpha_4$ ($+$ sign) or $\alpha_0-\alpha_4$ ($-$ sign).


\section{The Morrison--Park threefold} 
\label{sec:MP}

We are finally ready to explain how the construction of families of versally deformed, partially resolved ALE surfaces of type $D_4$ relates to the Morrison--Park threefold \cite{morrison-park}, engineering a generic F-theory model with matter of charge one and two.

\subsection{\texorpdfstring{$\U(1)$ gauge symmetry from geometry}{U1 gauge symmetry from geometry}}

In M-theory, some of the $\U(1)$ gauge symmetries come from the reduction of the three-form $C_3$ along harmonic two-forms of the compactification threefold $X_3$:
\begin{equation}\label{C3expansion}
  C_3 = \sum_I A_\mu^Idx^\mu \wedge D_I\ ,
\end{equation}
where $D_I$ are (the Poincar\'e duals of) divisors of $X_3$. 

Let us consider an elliptic fibration $X_3$. In the F-theory limit of M-theory, to have a massless $\U(1)$ gauge boson the elliptic fibration must have an extra section, or better a divisor that intersects the fiber at one point. We call this divisor $\mathcal{D}$. 
By combining the divisor $\mathcal{D}$ with the zero section $Z$ of the elliptic fibration and a proper vertical divisor $D_{\rm rest}$, one finds the $\U(1)$ generator \cite{grimm-timo}, i.e.\ the divisor that appears in $C_3 \sim A_\mu dx^\mu \wedge\omega_{\mathcal{D}}$:
\begin{equation}
  \omega_{\mathcal{D}} = \mathcal{D} - Z + D_{\rm rest}\ .
\end{equation}
The requirement that the elliptic fibration has an extra (rational or irrational) section leads to a specific form of the Weierstrass model, known as the \emph{Morrison--Park} (MP) \emph{threefold} \cite{morrison-park}:
\begin{equation}\label{EqMP}
W_\mathrm{MP}: \ \mathtt{y}^2 = \mathtt{x}^3+c_2 \mathtt{x}^2 + \left( c_1c_3-b^2c_0 \right) \mathtt{x} + c_0c_3^2-b^2c_0c_2+\frac{1}{4}b^2c_1^2\ ,
\end{equation}
where we have set $\mathtt{z}= 1$ (in the $\mathtt{z} \neq 0$ patch of the $\pp^{231}$ fiber). To obtain the standard  Weierstrass form $Y^2=X^3+f\,X+g$, one simply takes $\mathtt{y}=Y$ and $\mathtt{x} = X-\frac{c_2}{3}$.

As already mentioned, the geometry has been constructed so to have (at least) rank-one Mordell--Weil group; its generator is given by:
\begin{equation}\label{eq:Qmpa}
Q_\text{MP}:\left[ \mathtt{x}: \mathtt{y}: \mathtt{z}\right] = \left[c_3^2 -b^2 c_2 : -c_3^3+b^2 c_2 c_3 -\frac{1}{2} b^4 c_1 : b\right] \in \pp^{231}\ .
\end{equation}
The zero section sits at $Z: \left[\mathtt{x}:\mathtt{y}: \mathtt{z}\right]=\left[1:1:0\right]$. The non-Cartier divisor associated with the extra rational point is given by the equations
\begin{equation}\label{eq:divmpa}
\mathcal{D}_\text{MP} : \left\{ \mathtt{x} = \left(\frac{c_3^2}{b^2} - c_2\right)  ,\,\, \mathtt{y}= \left( -\frac{c_3^3}{b^3}+ \frac{c_2 c_3}{b} -\frac{1}{2} b c_1 \right) \right\} \subset W_\mathrm{MP}\ ,
\end{equation}
at least locally in the patch $\mathtt{z}=b \neq 0$. The MP geometry  admits a small, K\"ahler resolution, and we expect that the U(1) gauge boson remains massless at strong coupling \cite{bcv-fate}.

The MP threefold has two loci of point-like singularities, given by the following ideals:
\begin{equation}\label{eq:firstsingcurveX}
(\,\, \mathtt{y},\,\, \mathtt{x},\,\, b,\,\, c_3 \,\,)\ ,
\end{equation}
and 
\begin{equation}\label{eq:secondsingcurveX}
\left(\,\, \mathtt{y}, \,\,\, \text{minors}_{2\times 2} \begin{bmatrix}
b^2 & -c_3 & \mathtt{x} \\ -c_3 & \mathtt{x}+c_2 & \frac{c_1}{2} \\ \mathtt{x} & \frac{c_1}{2} & c_0 \end{bmatrix} \,\, \right) \ .
\end{equation}
The last one is not a complete intersection in the ambient space.
The extra divisor $\mathcal{D}_\text{MP}$ passes through both singular loci.

 In M-theory geometric engineering, these singularities correspond to massless states, given by M2-branes wrapping the shrunk $\mathbb{P}^1$, that are charged under the $\U(1)$ symmetry generated by the new divisor $\mathcal{D}_\text{MP}$. The actual charge is given by the intersection number of $\mathcal{D}_\text{MP}$ with the exceptional $\mathbb{P}^1$. This is due to the coupling of the M2-branes to the M-theory three-form $C_3\sim A_\mu dx^\mu \wedge \omega_{\mathcal{D}}$:
\begin{equation}
\int_\text{M2}C_3 =  \int A_\mu dx^\mu \int_{\mathbb{P}^1} \omega_{\mathcal{D}} = (\mathbb{P}^1\cdot \mathcal{D}_\text{MP})\int A_\mu dx^\mu\ .
\end{equation}
In the MP geometry, we have $\mathbb{P}^1\cdot \mathcal{D}_\text{MP}=2$ for the locus \eqref{eq:firstsingcurveX} and $\mathbb{P}^1\cdot \mathcal{D}_\text{MP}=1$ for the locus \eqref{eq:secondsingcurveX}. Therefore there are states with charge two and states with charge one.

\subsection{Morrison--Park is the universal flop of length two}
\label{sec:MPflop}

We will now show that the MP threefold \eqref{EqMP} is a particular threefold slice of the universal flop of length two (a sixfold). 
Indeed it can be obtained from \eqref{deformedD4amputquiv}, after identifying $x=\mathtt{y}$ and $y=\mathtt{x}$, by imposing the following equations:
\begin{equation}\label{eq:univtompaTs}
z = c_3(\zeta)  \ , \quad T_0 = \frac12  b(\zeta) \ , \quad T_1= \mathtt{x} + c_2(\zeta) \, \quad    T_2=c_0(\zeta) \ , \quad  T_3=c_1(\zeta) \ .
\end{equation}
The $\zeta_i$ are local coordinates on the twofold F-theory base (whereas $\mathtt{x}$ and $\mathtt{y}$ are coordinates along the fiber). In the patch where $z$ and $T_0$ can be taken as local coordinates,\footnote{This is the relevant patch to study the singularities of MP.} the last three equations give a cut of the universal flop sixfold. We also notice that the singular locus \eqref{eq:firstsingcurveX}, where charge-two states live, corresponds to the singular locus \eqref{SingLocCh2} of the universal flop sixfold, where the exceptional $\pp^1$ is quadratically embedded in a $\pp^2$. Analogously, the charge-one locus \eqref{eq:secondsingcurveX} corresponds to \eqref{SingLocCh1} in the sixfold, where the exceptional $\pp^1$ is an ordinary one.

The universal flop of length two can be written in the simpler form \cite{curto-morrison}
\begin{equation}\label{eq:univ}
W_\text{univ}:= x^2 + u y^2 +2 v y z + w z^2+(u w-v^2) t^2 = 0 \subset \cc^7_{(x,y,z,t,u,v,w)} \ 
\end{equation}
by applying to  \eqref{eq:D4verdef}  the following change of variables:
\begin{equation}
(T_0,T_1,T_2,T_3)=\left(\frac{t}{2},-u,-w,2v+y+z-u-w+\frac{t^2}{4} \right)\ .
\end{equation}
In this new form, the MP threefold is given, 
after identifying $x=\mathtt{y}$ and $y=\mathtt{x}$, by
\begin{equation}\label{eq:univtompa}
t= b(\zeta) \ , \quad z = c_3(\zeta)  \ , \quad u= - y -c_2(\zeta) \ ,\quad  v= -\frac{c_1(\zeta)}{2} \ , \quad  w=-c_0(\zeta) \ ,
\end{equation}
again with $\zeta_i$ coordinates on the F-theory twofold base.

\subsection{\texorpdfstring{Matrix factorizations and $\U(1)$ divisor}{Matrix factorizations and U(1) divisor}}
\label{sub:mf}

Call $R:=\cc[x,y,z,t,u,v,w]/(W_\text{univ})$ the coordinate ring of $W_\text{univ}$.
In \cite{curto-morrison} a $4 \times 4$ matrix factorization $(\Phi_\text{univ},\Psi_\text{univ})$ of \eqref{eq:univ} was presented, such that $M := \coker \Psi_\text{univ}$ is a so-called Cohen--Macaulay (CM) $R$-module (see \cite{collinucci-fazzi-valandro} for the relevant terminology.)
Thanks to \eqref{eq:univtompa}, the MP threefold \eqref{EqMP} will also admit a $4 \times 4$ matrix factorization (MF) $(\Phi_\text{MP},\Psi_\text{MP})$, see \eqref{eq:MFmpa}. By construction, these two matrices satisfy
\begin{equation}\label{eq:MFeqmpanoz}
\Phi_\text{MP}\cdot  \Psi_\text{MP} = \Psi_\mathrm{MP}\cdot  \Phi_\mathrm{MP} = W_\text{MP} \, 1_{4\times 4}\ .
\end{equation}
The MP threefold is a determinantal variety, since $W_\text{MP}= \text{Pfaff}\, \Phi_\text{MP}$. Over the hypersurface $W_\text{MP}=0$, the $4\times 4$ matrix $\Phi_\text{MP}$ has rank two. The singularities occur over the subloci where the rank of  $\Phi_\text{MP}$ is lower than two. There are two types of codimension-two loci where this happens; not surprisingly these correspond to \eqref{eq:firstsingcurveX} and \eqref{eq:secondsingcurveX}, upon using the equations \eqref{eq:univtompa}. The rank drops to zero at the charge-two locus \eqref{eq:firstsingcurveX}, while it drops to one at the charge-one locus \eqref{eq:secondsingcurveX}.
\begin{subequations}\label{eq:MFmpa}
\begin{align}
& \Psi_\text{MP}  = \begin{bmatrix}  \mathtt{y}-\frac{1}{2}b c_1 & -\mathtt{x} & -c_3 & -b \\  -c_1c_3-\mathtt{x} (c_2+\mathtt{x}) & \mathtt{y}+\frac{1}{2}b c_1 & -b (c_2+\mathtt{x}) & -c_3 \\  -c_0 c_3 & b c_0 & \mathtt{y}-\frac{1}{2}b c_1 & \mathtt{x} \\  bc_0 (c_2+\mathtt{x}) & -c_0 c_3 & c_1 c_3+\mathtt{x} (c_2+\mathtt{x}) & \mathtt{y}+\frac{1}{2}b c_1 
\end{bmatrix} \ , \\
&\Phi_\text{MP}  = \begin{bmatrix}  \mathtt{y}+\frac{1}{2}b c_1 & \mathtt{x} & c_3 & b \\  c_1c_3+\mathtt{x} (c_2+\mathtt{x}) & \mathtt{y}-\frac{1}{2}b c_1 & b (c_2+\mathtt{x}) & c_3 \\  c_0 c_3 & -b c_0 & \mathtt{y}+\frac{1}{2}b c_1 & -\mathtt{x} \\  -bc_0 (c_2+\mathtt{x}) & c_0 c_3 & -c_1 c_3-\mathtt{x} (c_2+\mathtt{x}) & \mathtt{y}-\frac{1}{2}b c_1 
\end{bmatrix} \ .
\end{align}
\end{subequations}
(The MF \eqref{eq:MFmpa} can be straightforwardly completed to a globally-defined one in $\pp^{231}[\mathtt{x}:\mathtt{y}:\mathtt{z}]$.)

In \cite{collinucci-fazzi-valandro} it was shown how a $4\times 4$ MF of a threefold can be associated with a family of non-Cartier divisors intersecting the exceptional locus: As we have said, the MF defines the CM module $M := \coker \Psi_\text{univ}$. 
The rank-two module $M$ becomes a locally-free sheaf in the resolved threefold, i.e.\ a rank-two vector bundle $\cV$ generated by its sections. The Poincar\'e dual of the first Chern class of the vector bundle is a divisor $\mathcal{D}$. It is related to the $\U(1)$ gauge boson in the M-theory compactification. The divisor is given by the locus where two generic sections of the rank-two vector bundle become parallel. This locus can be identified already in the singular space, by requiring that two sections of $M$ be proportional to each other. 
To do this we use the isomorphism between $\coker \Psi_\text{MP}$ and $\im \Phi_\text{MP}$ explained in \cite{collinucci-fazzi-valandro}: When the domain of the map $\Phi_\text{MP}$ is restricted to be $\coker \Psi_\text{MP}$, the map is bijective (this is valid on generic points of the MP threefold).
Hence, the locus where two sections of $\coker \Psi_\text{MP}$ are parallel is the same as the locus where two sections of $\im \Phi_\text{MP}$ are parallel.
Since the image $\im \Phi_\text{MP}$ is generated by the columns of $\Phi_\text{MP}$, we can choose two columns of $\Phi_\text{MP}$ and find the locus where these become parallel. 

Take e.g. the last two columns of $\Phi_\text{MP}$. The locus we are looking for is given by the condition
\begin{equation}
  \mbox{rank} \,\begin{bmatrix}
  c_3& b \\  b(c_2+\mathtt{x}) & c_3 \\ \mathtt{y}+\tfrac12 bc_1 & -\mathtt{x} \\  -c_1c_3-\mathtt{x}(c_2+\mathtt{x}) & \mathtt{y}-\tfrac12 bc_1 \\ \end{bmatrix} \leq 1\ .
\end{equation}
When all the $2\times 2$ minors vanish, we obtain the vanishing locus of the following ideal:
\begin{equation}\label{EgIdealDiv}
\left( b^2(c_2+\mathtt{x}) -c_3^2, \mathtt{y}b +c_3 \mathtt{x} +\tfrac{1}{2}b^2 c_1, c_3 \mathtt{y} +b\left( \tfrac{1}{2} c_1c_3 +\mathtt{x}(\mathtt{x}+c_2)\right), \eqref{EqMP}\right) \ .
\end{equation}
We can make independent choices by taking  e.g. the second and last columns or the first and last.
Taking generic combinations of columns and requiring them to be parallel gives a whole family of Weil divisors \cite{collinucci-fazzi-valandro}, of the form
\begin{equation}\label{eq:Z+k}
M_{\mathcal{D}} \cdot \begin{pmatrix} k_1 \\ k_2 \\ k_3 \end{pmatrix}= 0\ ,
\end{equation}
with 
\begin{align}
M_{\mathcal{D}} = \begin{bmatrix}
 b^2 (c_2+\mathtt{x})-c_3^2 & \frac{1}{2}c_1 b^2+\mathtt{y} b+c_3 \mathtt{x} & - b (\tfrac{1}{2}c_1 c_3+ \mathtt{x} (c_2+\mathtt{x}))-c_3 \mathtt{y} \\
 \frac{1}{2}c_1 b^2-\mathtt{y} b+c_3 \mathtt{x} & b^2 c_0-\mathtt{x}^2 & \mathtt{x} \mathtt{y}-\frac{1}{2} b (2 c_0 c_3+c_1 \mathtt{x}) \\
 b \left(\mathtt{x} (c_2+\mathtt{x})-\frac{1}{2}c_1 c_3\right)-c_3 \mathtt{y} & b c_0 c_3+\mathtt{x} \left(\frac{1}{2}b c_1+\mathtt{y}\right) & -(\frac{1}{2}b c_1+ \mathtt{y})^2-b^2 c_0 (c_2+\mathtt{x}) \end{bmatrix}\ .
\end{align}
Taking e.g. $\vec{k}=(1,0,0)^\text{t}$ gives the ideal \eqref{EgIdealDiv}, corresponding to the non-Cartier divisor $\mathcal{D}_\text{MP}$ found in \cite{morrison-park} and defined in \eqref{eq:divmpa}. However notice that the former is just \emph{one} representative in a whole family the MF is capable of providing us.

The generic choice in the family can also be constructed in the universal flop sixfold directly, and upon using \eqref{eq:divmpa} it matches with \eqref{eq:Z+k}. By using the techniques discussed in \cite{collinucci-fazzi-valandro}, one can show that it intersects once the exceptional fiber at the origin of the sixfold. In the language of section~\ref{sec:D4-quiv}, this locus corresponds to $t_1=...=t_4=0$ (i.e.\ to case \ref{locus4res}). The exceptional $\mathbb{P}^1$ is the sphere $\alpha_4$ in the ALE fiber. Now consider the exceptional $\mathbb{P}^1$ on top of \eqref{eq:firstsingcurveX}; it is in the class $\alpha_0$, and when intersected with \eqref{eq:secondsingcurveX} it splits into two $\mathbb{P}^1$'s that coincide at the origin of the sixfold. Hence the divisor we have constructed will intersect twice the exceptional $\mathbb{P}^1$ at the locus \eqref{eq:firstsingcurveX} where the matrix rank drops to zero, while it will intersect once the exceptional $\mathbb{P}^1$ at the locus \eqref{eq:secondsingcurveX} where the matrix rank drops to one.

\subsection{Laufer threefold}
\label{sub:laufer}

In \cite{collinucci-fazzi-valandro}, we discussed in full detail the Laufer threefold \cite{aspinwall-morrison}, which is described by the following equation:\footnote{This is Laufer's threefold for $n=1$; see \cite[Eq. (69)]{aspinwall-morrison}.}
\begin{equation}
x^2+y^3-tz^2-yt^3 = 0\ .
\end{equation}
This is also a threefold cut of  the universal flop of length two (and also a specialization of MP). It is obtained from \eqref{deformedD4amputquiv} by imposing
\begin{equation}\label{eq:univtolauf}
T_0= \frac{t}{2} \ , \quad T_1 = -y  \ ,\quad T_2= t \ ,\quad  T_3= z+t+\frac{t^2}{4} \ .
\end{equation}
The resulting threefold is singular at $x=y=z=t=0$. Notice that at this point all the $T_i$'s vanish. In terms of the covering variables $t_i$, this means $t_1=t_2=t_3=t_4=0$, i.e.\ we are in the case \ref{locus4res} of sections~\ref{subsec:DefK3} and \ref{subsec:ResK3}. The ALE space develops a $D_4$ singularity, and we resolve its central node (the black one in figure \ref{fig:dynkins}) by a simultaneous resolution. 

This singular threefold with its two resolutions is the simplest example of \emph{length-two flop} (with the conifold flop being instead length-one). The exceptional $\pp^1$ in this case is `length-two', i.e.\ is an example of \emph{non-reduced scheme} (roughly, an algebraic variety defined by $w^2=0$, for some local coordinate $w$).\footnote{More precisely, the coordinate ring $R$ of the singular variety $X:=\text{Spec} R$ (i.e.\ the scheme) contains a nilpotent element. The length is simply the dimension of the (complex) vector space $\mathcal{O}_X(X)$. For a brief account on the subject see e.g. this \href{http://math.mit.edu/~mckernan/Teaching/09-10/Spring/18.726/l_6.pdf}{\nolinkurl{page}} or, for a more complete treatment, section 4.2 of Vakil's lectures \cite{FOAG}.} The $\pp^1$ is intersected once by the $\U(1)$ divisor. However, this two-cycle supports a bound state of two superposed membranes, which means that charge-two matter is also possible by this mechanism \cite[Sec. 6]{collinucci-fazzi-valandro}.\footnote{The original Laufer example \cite{laufer} was later generalized in \cite{pinkham,morrison}: These too are specializations \cite[Sec. 4.2]{aspinwall-morrison} of the universal flop by \cite[Thm. 3]{curto-morrison}, and their singular geometries host charge-two matter.}

\section{The general picture for other ALE spaces} 
\label{sec:gen-K3}

We have seen the connection between  the universal flop of length two and M-theory models with one massless $\U(1)$ and matter with charge one and two. The universal flop  sixfold has two loci of singularities that the threefold cut inherits. After resolution, the volume of the exceptional $\mathbb{P}^1$ over one locus is twice the volume of the exceptional $\mathbb{P}^1$ over the second locus. If the two loci intersect, the quadratically-embedded $\mathbb{P}^1$ splits into two $\mathbb{P}^1$. We have seen this both algebraically and by studying the ALE fiber.

The same analysis can be performed for models with length higher than two, which will produce matter with charge higher than two. As we have seen in section~\ref{sec:quivers}, also these $n$-folds are ALE fibrations over the space of deformation parameters.

\subsection{ALE fibrations with higher-charge states}

One can straightforwardly repeat the analysis of section~\ref{sec:D4-quiv} to cases corresponding to higher length. In order to get charge-three states one starts from the family of deformations of an ALE space with one $E_6$ singularity (i.e.\ the universal flop of length three), for charge four (length four) one studies the deformations of an $E_7$ singularity, whereas for charge five and six (length five and six) those of an $E_8$ singularity.

To illustrate how the technique works in higher charge, we will consider the model with maximal possible length. As we said, the starting point is an ALE space with an $E_8$ singularity. The $E_8$ root lattice is eight-dimensional, hence the holomorphic two-form has eight deformation parameters $t_1,\ldots,t_8$, corresponding to the (complex) volume of the simple roots. The highest root is (see the Dynkin diagram in figure \ref{fig:dynkins})
\begin{equation}
\alpha_0=2\alpha_1 +3\alpha_2 +4\alpha_3 +5\alpha_4 +3\alpha_5 +2\alpha_6 +4\alpha_7 +6\alpha_8\ ,
\end{equation}
with complex parameter $2t_1 +3t_2 +4t_3 +5t_4 +3t_5 +2t_6 +4t_7 +6t_8$.

In the length-two case, the crucial point to construct matter of charge two (in the M-theory threefold) was the simultaneous resolution of the simple root that appears with the highest weight in the highest root (i.e.\ the black node of the Dynkin, labeled by a $2$). Hence, in order to have charge six, we need to take the simultaneous resolution of the central root in the $E_8$ Dynkin diagram, i.e.\ 
\begin{equation}
J = {J'}^{E_8} + \xi \alpha^\ast_8\ .
\end{equation}
Now, consider the spheres that are integral  linear combinations of simple roots with non-zero coefficient along $\alpha_8$, i.e.\ $\alpha=\sum_{i=1}^8 n_i \alpha_i$ with $n_8\neq 0$ (and $\alpha^2=-2$).
Looking at the Dynkin diagram, one can see that all values of $n_8$ are possible from 1 to 6.
Take the family of ALE spaces with an $E_8$ singularity. The locus where one of these spheres shrinks is codimension-one in the parameter space spanned by $t_1,\ldots,t_8$. Over such loci the ALE fiber develops an $A_1$ singularity.
After resolution, the exceptional $\mathbb{P}^1$ has real volume $n_8\xi$.
For instance, along the locus $t_0=0$, the sphere in the class $\alpha_0$ shrinks, generating an $A_1$ singularity. After resolution, the exceptional $\mathbb{P}^1$ in the class $\alpha_0$ has real volume 
\begin{equation}
J\cdot \alpha_0 = 6\xi \ .
\end{equation}
Since $n_8=1,2,...,6$, \emph{all} charges in $\{1,2,...,6\}$ are present.

Notice that making the simultaneous resolution of the root corresponding to the node with label 5 in the $E_8$ Dynkin diagram corresponds to $J = {J'}^{E_8} + \xi \alpha^\ast_4$. The highest root has now real volume $J\cdot \alpha_0 = 5\xi$ (i.e.\ we have matter with charge up to five).

\subsection{Quivers for higher charge}
\label{sub:quivhigh}

The algebraic description of the models with high charge can be approached by using the quiver techniques.
In this section we will present the quivers constructed in \cite[Sec. 4]{karmazyn} whose gauge invariants satisfy a single relation that is equivalent to the versally deformed ADE singularity (admitting a simultaneous resolution of the black node, as per figure \ref{fig:dynkins}). The case of $D_4$ was already presented in section \ref{sec:quivers}, so we will neglect it in what follows.

\begin{figure}[ht!]
\centering
\includegraphics[scale=1.25]{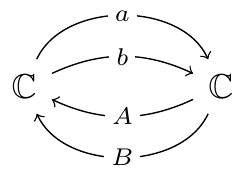}
\caption{The quiver with relations reproducing the threefold family of deformed $A_1$ surface singularities admitting a simultaneous resolution.}
\label{fig:length1}
\end{figure}

We start with $A_1$. The quiver is depicted in figure \ref{fig:length1}, and the relations are
\begin{equation}\label{eq:length1}
A_1: \quad aA-bB=T_0\ , \quad Bb-Aa=-T_0\ .
\end{equation}
The gauge invariants are the paths $x=aA, y=bA, z=aB$, satisfying $x^2=yz$ in the undeformed case (as can be checked by applying \eqref{eq:length1} with $T_0=0$), and $x^2=yz + T_0 x$ in the deformed one.

\begin{figure}[ht!]
\centering
\includegraphics[scale=1.25]{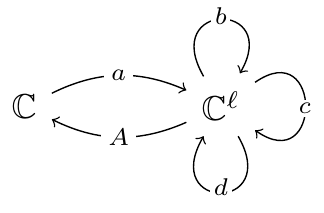}
\caption{The quiver with relations reproducing the threefold family of deformed $E_6,E_7,E_8^{(5)},E_8^{(6)}$ surface singularities admitting a simultaneous resolution, with $\ell=3,4,5,6$ respectively (equal to the label of the black node in the corresponding non-affine Dynkin of figure \ref{fig:dynkins}).}
\label{fig:length3}
\end{figure}

We now move to the cases $E_6, E_7, E_8^{(5)}, E_8^{(6)}$. Remember that $E_8$ has two possible colorings (see figure \ref{fig:dynkins}), corresponding to which $\pp^1$ we want to give a real volume in the homology of the partially resolved, versally deformed ALE spaces. The first (second) coloring corresponds to a $\pp^1$ associated with the node of multiplicity five (six), hence to (the presence of) a matter locus in the threefold with charge five (six). For all these cases the quiver is the one in figure \ref{fig:length3}.
However each case requires different relations among the arrows, as follows.
\begin{subequations}\label{eq:lengths}
\begin{equation}\label{eq:length3}
\begin{split}
E_6\ &\  (\ell=3)\ : \quad x:=ac^2bcA\ , \quad y:=ac^2A\ , \quad z:=acA\ , \quad x^2-z^2x+y^3= 0 \ ; \\ &\ aA=T_0^2-T_5  \ , \quad Aa=d^2-T_5 1_3 \ , \quad dA = AT_0  \ , \quad ad= T_0 a \ ,\\ &\ b^3-T_2 b -T_1 1_3 = 0\ , \quad c^3-T_4 c-T_3 1_3 =0 \ , \quad b+c+d - \frac{T_0}{3}1_3=0\ .
 \end{split}
\end{equation}
\begin{equation}\label{eq:length4}
\begin{split}
E_7\  &\ (\ell=4)\ : \quad x:=ac^3bc^2A\ , \quad y:=ac^3A\ , \quad z:=acA\ , \quad x^2-y^3+yz^3= 0 \ ; \\ &\ aA=T_0^3-T_2T_0-T_1 \ , \quad Aa=d^3-T_2d-T_1 1_4 \ , \\ &\ dA = AT_0  \ , \quad ad= T_0 a \ , \quad b^2-T_31_4 = 0\ , \\ &\ c^4-T_4 c^2-T_5c-T_6 1_4 =0 \ , \quad b+c+d - \frac{T_0}{4}1_4=0\ .
 \end{split}
\end{equation}
\begin{equation}\label{eq:length5}
\begin{split}
E_8^{(5)}\  &\ (\ell=5)\ : \quad x:=ac^3bc^2A\ , \quad y:=ac^3A\ , \quad z:=acA\ , \quad x^2-y^3-z^5= 0 \ ; \\ &\ aA=T_0^4-T_3T_0^2-T_2T_0-T_1 \ , \quad Aa=d^4-T_3d^2-T_2d-T_1 1_5 \ , \\& \ dA = AT_0 \ , \quad ad= T_0 a\ , \quad b-c + \frac{T_0}{5} 1_5=0 \\ &\ cbc+c^2b+cb^3+T_7cb+T_6c+T_41_5 =0 \ , \\ &\ (c+b^2)^2+bcb+T_7(c+b^2)+T_6b+T_5 1_5 = 0 \ .
 \end{split}
\end{equation}
\begin{equation}\label{eq:length6}
\begin{split}
E_8^{(6)}\  &\ (\ell=6)\ :  \quad x:=ac^2bc^2bcbc^2A\ , \quad y:=ac^2bc^2A\ , \quad z:=acA\ , \quad x^2-y^3+z^5= 0 \ ; \\ &\ aA=T_0^5-T_4T_0^3-T_3T_0^2-T_2T_0-T_1 \ , \\ &\ Aa=d^5-T_4d^3-T_3d^2-T_2d- T_1 1_6 \ , \\& \ dA = AT_0  \ , \quad ad= T_0 a\ , \quad b+c +d - \frac{T_0}{6}1_6=0 \\ &\ b^2 -T_7 1_6 =0 \ , \quad c^3-T_6 c-T_5 1_6 =0\ . \end{split}
\end{equation}
\end{subequations}
In general (i.e.\ with generic choices of parameters $T_i$), concrete examples will be extremely lengthy, as we discuss in the next subsections. However one should bear in mind that, ultimately, all cases can be brought to the following forms through coordinate redefinitions:
\begin{subequations}
\begin{align}\label{eq:Eelliptic}
\begin{split}
E_6 {}: Y^2 -Y Z^2 =&\ X^3+ \epsilon_2(T_i) X Z^2+ \epsilon_5(T_i) X Z + \epsilon_6(T_i) Z^2\, +\\&+\epsilon_8(T_i) X+\epsilon_9(T_i) Z+\epsilon_{12}(T_i) \ ;
\end{split}\\ \cr
\begin{split}
E_7 {}: Y^2=&\ X^3+16 X Z^3 + \epsilon_2(T_i) X^2 Z+ \epsilon_6(T_i) X^2+ \epsilon_8(T_i) X Z\, +\\& +\epsilon_{10}(T_i) Z^2+\epsilon_{12}(T_i) X+\epsilon_{14}(T_i) Z+\epsilon_{18}(T_i)  \ ; 
\end{split}\\ \cr
\begin{split}
E_8^{(5)} \ , \ E_8^{(6)} {}: Y^2=&\ X^3-Z^5 + \epsilon_2(T_i) X Z^3+ \epsilon_8(T_i) X Z^2+ \epsilon_{12}(T_i) Z^3\, +\\& +\epsilon_{14}(T_i) X Z+\epsilon_{18}(T_i) Z^2+\epsilon_{20}(T_i) X+\epsilon_{24}(T_i) Z+\epsilon_{30}(T_i) \ .
\end{split}
\end{align}
\end{subequations}
As the reader can appreciate, all these cases can be regarded as patches of elliptic fibrations, and it is reasonable to suppose that there is a model fibered by ALG spaces as well. Here, the $T_i$'s will be appropriately covariant with respect to the complementary subgroup of the full Weyl group of $E_n$ (as per figure \ref{fig:dynkins}), with the $\epsilon_j$'s being functions of the elementary symmetric polynomials of degree $j$ in the $T_i$'s \cite{katz-morrison}. In order to make a CY threefold, one shall regard the $T_i$'s as sections of an appropriate power of the canonical bundle of the base of the fibration.

\section{Conclusions} 
\label{sec:conc}

\subsubsection*{Building threefolds}

In this paper, we introduced a `natural' class of models for M-theory (F-theory) compactifications to five (six) dimensions, that contain an Abelian vector multiplet, and hypers of charges ranging from one to six.
The attribute `natural' is left deliberately vague here. There are other proposals that can realize charges in F-theory higher than six \cite{taylor-turner,raghuram-taylor} that clearly do not fall into this class. 


The models we present here become inhumanely lengthy, when written in full generality. It would be desirable to specialize to a few interesting, but more succinct examples. This should be achievable by means of the algorithm we presented in appendix~\ref{app:explicit}.

The models of charge $3, 4, 5$ and $6$ are all derived as families of $E_n$-type ALE spaces, which are naturally elliptic and in `Tate form'. However, other possible slicings of these families might be possible. For instance, for the charge-two case, we were able to reproduce the Morrison--Park model \cite{morrison-park} by taking a slice that is not vertical with respect to the family of ALE fibers, but cuts through the family mixing fiber and $T_i$ directions.

Finally, we remark that all these high-charge models can easily be used to construct CY fourfolds (or even fivefolds and sixfolds).

\subsubsection*{Geometric transitions}

It would be interesting to see whether `geometric transitions'\footnote{For a review from both the mathematics and the physics perspectives see \cite{Rossi:2004eq,Acharya:2004qe}.} -- akin to the well-known one for the conifold \cite{Candelas:1989ug,Strominger:1995cz,Greene:1995hu} -- can be established for (threefold slices of) the higher-length flops. These are topology-changing transitions, whereby one passes from the resolution of the singularity (containing a single $\pp^1$, i.e. a holomorphic two-sphere, in the case of the conifold) to its complex deformation (with a single three-sphere for the conifold). In general, the Milnor number of a singularity gives the number of three-spheres in the deformation. This number can easily be computed for threefold slices of the universal flops which contain an ADE singularity. 

It would be interesting to furnish a gauge theory interpretation of the transition in terms of its low-energy dynamics.

\subsubsection*{\boldmath$\U(1)$ divisors}
The second main result of this paper is the application of the theory of Eisenbud's matrix factorizations to single out the family of $\U(1)$ divisors in the compactification. Usually, in an F-theory situation, one seeks out the extra sections of the elliptic fibration to construct a homology class upon which to reduce the supergravity three-form potential.
By using matrix factorizations, we can actually find the whole linear system of divisors with which to construct it. The method does not require any map to some birational non-Weierstrass model, but is naturally derived from the inherent ALE-fibration structure of the family.

The matter charges can essentially be read off of the size of the matrix factorization of the singularity, which in our examples can be explicitly constructed with the method explained in appendix~\ref{app:explicit}. The maximum allowed charge is then $\ell$ for a $2\ell \times 2\ell$ matrix pair, with the corresponding charged hypers sitting at the locus where the matrix rank drops to zero. Generically, there will be other $\ell-1$ special singular loci where the rank drops to $r$, with $r$ ranging from $1$ to $\ell-1$. In analogy with the charge-two case of section~\ref{sub:mf}, we claim that these loci correspond to the lower-charge states. Hence, our models typically realize all the charges from 1 (where the rank drops to $\ell-1$) to $\ell$ (where the rank drops to zero). This gives a rather practical method to single out the various charged states.

\subsubsection*{Elliptic fibrations and ALG fibrations}
Most of the analysis of this paper was done in algebraic terms, with the results cast in terms of ALE-fibrations (using Kronheimer's identification of the algebraic and differentio-geometric descriptions of such spaces \cite{kronheimer-ALE}).  However, we noticed in a few places that the algebraic description contained the structure of an elliptic fibration; it would be interesting to know if we could replace that with an ALG fibration (perhaps using recent results of Hein, Chen, and Chen \cite{MR2869021,arXiv:1505.01790,arXiv:1508.07908,arXiv:1603.08465}).  If so, we would have even greater confidence in the applicability of these results to F-theory, since an appropriate metric on the space would have been given.


\section*{Acknowledgments} 

We have benefited from discussions with F.~Apruzzi, C.~Beil, D.~Berenstein, O.~Bergman, J.~Halverson, J.~J.~Heckman, J.~Karmazyn, S.~Schäfer-Nameki, and W.~Taylor. M.F.~is particularly indebted to J.~Karmazyn, for numerous enlightening explanations and for sharing a draft of \cite{karmazyn} prior to publication. The work of M.F.~and D.R.M.~was performed in part at the Simons Center for Geometry and Physics, Stony Brook University during the XVI Simons Summer Workshop. The work of A.C.,~M.F., and D.R.M.~was performed in part at the Banff International Research Station and at the Aspen Center for Physics, which is supported by the NSF grant PHY-1607611. M.F.~wishes to thank the Weizmann Institute of Science for hospitality during the final stages of this work, and acknowledges financial support from the Aspen Center for Physics through a Jacob Shaham Fellowship Fund gift. A.C.~is a Research Associate of the Fonds de la Recherche Scientifique F.N.R.S.~(Belgium). The work of A.C.~is partially supported by IISN - Belgium (convention 4.4503.15), and supported by the Fonds de la Recherche Scientifique - F.N.R.S.~under Grant CDR J.0181.18. The work of M.F.~is partially supported by the Israel Science Foundation under Grant No.~504/13,~1696/15,~and 1390/17, and by the I-CORE Program of the Planning and Budgeting Committee. D.R.M.~wishes to thank the Kavli Institute for the Physics and Mathematics of the Universe at the University of Tokyo for hospitality during the final stages of this work, and acknowledges financial support from NSF grants PHY-1307514, 
DMS-1440140 (at the Mathematical Sciences Research Institute, Berkeley),
PHY-1620842, and 
PHY-1748958 (at the Kavli Institute for Theoretical Physics, U.C. Santa Barbara), and from a Fellowship in Theoretical Physics from the Simons Foundation 
[award \#562580]. The work of R.V.~is partially supported by ``Fondo per la Ricerca di Ateneo - FRA 2018'' (UniTS).


\newpage

\appendix

\section{Producing explicit examples with higher charge}
\label{app:explicit}


In each ADE case the versal deformation can be calculated with the aid of a simple computer algebra code, that we reproduce below.\footnote{We would like to thank J.~Karmazyn for help with the code. A slightly different version of the code is also provided in \cite{karmazyn}.} E.g. for the versal deformation of $D_4$ one types in Magma \cite{magma}:\footnote{For short calculations (of at most $120$ seconds) one can use the online Magma calculator available at this \href{http://magma.maths.usyd.edu.au/calc/}{\nolinkurl{page}}.}
\begin{verbatim}
K := RationalField();
Kt<T0, T1, T2, T3, z, y> := RationalFunctionField(K,6); 
F<a, A, x, d, c, b> := FreeAlgebra(Kt, 6);
B := [ 
z-a*b*A, 
y-a*c*A, 
x-a*b*c*A,
a*A-2*T0, b*b-T1, c*c-T2, d*d-T3, 
A*a+b+c+d-T0
];
G:=GroebnerBasis(B,6);

P:=0;
f:=a*b*c*(T0-b-c-d) - (1/2)*P*a;

g1:=NormalForm(f,G);
g2:=NormalForm(g1*b,G);  
g3:=NormalForm(g2*c,G); 
g4:=NormalForm(g3*A-(1/2)*P*x+(1/4)*P*P,G); 

printf"Charge-2 example with T0, T1, T2, T3 generic deformation 
parameters (i.e.\ universal flopping algebra of length 2)
after base change "; 

printf"\n \n";

printf"Hypersurface equation is 0 = - x^2 + "; g4; printf"\n \n";

printf"The polynomial P is "; P; printf"\n \n";
\end{verbatim}
Notice that in the definition of $\mathtt{f}$, which is simply $xa-\tfrac{1}{2}Pa$, we are using the relation $Aa=-b-c-d+T_0$. This should be modified in the appropriate way for all other cases, as explained below.

The relations and gauge invariants $x,y,z$ in the definition of $\mathtt{B}$ are taken from \eqref{eq:rellength2} and \eqref{eq:D4gaugeinv} respectively.  (It is important to input the various factors in each monomial of the relations in the order specified in \eqref{eq:lengths}. This is because the code treats the relations as a non-Abelian algebra.) The output is precisely \eqref{eq:D4verdef}, which maps to the form \eqref{eq:D4fulldef} (useful for directly selecting which homology two-sphere should have non-zero volume) under base change \eqref{eq:baseD4}. 

The above output contains a linear term in $x$, which is not present in the form \eqref{eq:D4fulldefweyl}. We can easily cancel it by completing the square.
\begin{equation}
-x^2 - P(y,z,T_0,\ldots,T_3)x + Q(y,z,T_0,\ldots,T_3) =0 \ \xrightarrow{x \to x-\tfrac{P}{2}} -x^2 + \frac{1}{4}P^2 + Q =0\ .
\end{equation}
To achieve this, one steps the same code, but now replaces the right-hand side of $\mathtt{P}:=0;$ with the coefficient of the term in $\mathsf{x}$.\footnote{This coefficient can be calculated with the command $\mathtt{P:=MonomialCoefficient(NormalForm(x*x,G),x);}$ upon removing $\mathsf{x}$ from the definition of $\mathtt{F}$ (and decreasing the order of the $\mathtt{FreeAlgebra}$ by one) and the relation $\mathtt{x}-\mathtt{a}\ldots\mathtt{A}$ from $\mathtt{B}$.} Explicitly:
\begin{verbatim}
K := RationalField();
Kt<T0, T1, T2, T3, z, y>:=RationalFunctionField(K,6); 
F<a, A, x, d, c, b> := FreeAlgebra(Kt, 6);
B := [ 
z-a*b*A, 
y-a*c*A, 
x-a*b*c*A,
a*A-2*T0, b*b-T1, c*c-T2, d*d-T3, 
A*a+b+c+d-T0
];
G:=GroebnerBasis(B,6);

P:=(-2*T0^3 - 2*T0*T1 - 2*T0*T2 + 2*T0*T3 - 2*T0*z - 2*T0*y);
f:=a*b*c*(T0-b-c-d) - (1/2)*P*a;

g1:=NormalForm(f,G);
g2:=NormalForm(g1*b,G);  
g3:=NormalForm(g2*c,G); 
g4:=NormalForm(g3*A-(1/2)*P*x+(1/4)*P*P,G); 

printf"Charge-2 example with T0, T1, T2, T3 generic deformation
parameters (i.e.\ universal flopping algebra of length 2)
after base change "; 

printf"\n \n";

printf"Hypersurface equation is 0 = - x^2 + "; g4; printf"\n \n";

printf"The polynomial P is "; P; printf"\n \n";
\end{verbatim}
Finally, with little effort we can bring the versally deformed $D_4$ surface into elliptic form (and presumably describe it in terms of an ALG-fibration), which is particularly useful for F-theory applications. The output of the second code (neglecting the $-x^2$ term) is a polynomial of the form,
\begin{equation}
\alpha z^2 + \beta z + (\gamma + y + z)yz+\delta y + \epsilon y^2  + \kappa\ ,
\end{equation}
with $\alpha,\ldots,\kappa$ depending only on the $T_i$'s. (This could now be brought into form \eqref{eq:D4fulldefweyl} via a simple coordinate redefinition.) We can bring the above into the standard form $z^3 + fz + g$ via a (linear) change of variables.\footnote{The same arguments would apply to $y$ verbatim, given the symmetry between the latter and $z$ in the above polynomial.} Indeed, shifting $y \to y + \phi z$ makes a cubic term in $z$ appear, with coefficient $\phi(1+\phi)$. The coefficient of the quadratic term is instead $b_2:=\alpha +\phi  (\gamma +\epsilon  \phi )+y(1+2 \phi)$. Now shift $z \to z - \frac{1}{3\phi (1+\phi)}b_2$. This produces a polynomial
\begin{equation}\label{eq:D4elliptic}
\phi(1+\phi)\,z^3 + \frac{1}{3\phi(1+\phi)}f(y,T_i)\, z + \frac{1}{(3\phi(1+\phi))^3}g(y,T_i) \ ,
\end{equation}
with $f$ and $g$ depending on $y,T_0,\ldots,T_3$. By appropriately rescaling $y$ and $z$, we can cancel the unpleasant factors in $\phi$, and we are done.\footnote{Alternatively, a different Ansatz $y \to \psi y + \phi z$, with a judicious choice of $\psi,\phi$, can directly produce the sought form $z^3 +f(y,T_i)z +g(y,T_i)$. Notice that in \eqref{eq:D4elliptic} the roles of the ``fiber coordinates'' $x,y$ in standard F-theory notation are played by our $z$ and $x$ respectively.}

\subsubsection*{The simplest case of $A_1$}

In the simplest of all cases, namely $A_1$, corresponding to the conifold threefold with a charge-one matter locus only, one inputs the simplified code:
\begin{verbatim}
K := RationalField(); 
Kt<T0, z, y>:=RationalFunctionField(K,3); 
F<a, A, b, B, x> := FreeAlgebra(Kt, 5);
D := [ 
x-a*A, 
y-b*A,
z-a*B,
a*A-b*B-T0,
B*b-A*a+T0
];
G:=GroebnerBasis(D,6);

P:=0;
f:=a*A*a - (1/2)*P*a;
g:=NormalForm(f*A-(1/2)*P*x+(1/4)*P*P,G);

printf"Hypersurface equation is 0 = - x^2 +"; g; printf"\n \n";

printf"The polynomial P is "; P; printf"\n \n";
\end{verbatim}
Doing so, one discovers that the coefficient of the linear term in $\mathtt{x}$ is precisely $\mathtt{T0}$. Stepping the modified code obtained by replacing $\mathtt{P:=0;}$ with $\mathtt{P:=T0;}$ one obtains
\begin{equation}\label{eq:A1verdef}
-x^2+zy+\frac{1}{4}T_0^2=0\ ,
\end{equation}
which is precisely \eqref{eq:A1coni} after base change $\alpha \mapsto \tfrac{T_0^2}{4}$ to a Weyl-invariant coordinate.\footnote{An obvious coordinate redefinition is also needed.}

\subsubsection*{Tuning a non-generic matter locus}

The above procedure for $D_4$ produces an explicit albeit generic example with a locus of charge two, and a locus of charge one. One can also greatly simplify the polynomials $f,g$ by selecting nongeneric values for the deformation parameters. The choice should be dictated by the logic explained in section \ref{sec:D4-quiv}. If one wants to tune a specific matter locus on the fourfold base of the sixfold, corresponding to having only a certain two-sphere (or linear combination of spheres) acquire non-zero volume in the homology of the ALE space, one should impose those conditions at the level of the deformation parameters inside the above code. One should simply remember to lower the dimension of
\begin{verbatim}
Kt<T0, T1, T2, T3, z, y>:=RationalFunctionField(K,6)
\end{verbatim}
($6$ in the generic case), if one fixes a relation among some of the $T_i$'s.

\subsubsection*{Examples with higher charge}
The above code can easily be adapted to all universal flopping algebras of length $\ell=3,\ldots,6$ (with $\ell=2$ corresponding to the $D_4$ case). First off, one should modify the line
\begin{verbatim}
Kt<T0, T1, T2, T3, z, y>:=RationalFunctionField(K,6)
\end{verbatim}
to include up to $T_7$ (thereby appropriately modifying the dimension of $\mathtt{Kt}$). Then one should input in $\mathtt{B}$ the correct relations and gauge invariants, taken from \eqref{eq:length3}, \eqref{eq:length4}, \eqref{eq:length5}, \eqref{eq:length6} for $E_6,E_7,E_8^{(5)}, E_8^{(6)}$ (i.e.\ charge three, four, five, six) respectively. Moreover, for high charge examples it might be necessary to increase the number of parameters in the calculation of a Gröbner basis for $\mathtt{B}$, by simply replacing the $6$ in the line below by some higher number (starting from $8$, and gradually increasing it as needed):
\begin{verbatim}
G:=GroebnerBasis(B,6);
\end{verbatim}
Finally, one should input one polynomial $\mathtt{g}$ per arrow in the definition of the gauge invariant $x=a \cdots A :\cc \to \cc$ as quiver path, with $\mathtt{g1}$ and $\mathtt{g}_\text{last}$ corresponding to $a$ and $A$ respectively. E.g. for $E_6$ we have $x=ac^2bcA$, hence we should input the correct $\mathtt{f}$ and 
\begin{verbatim}
g1:=NormalForm(f,G);
g2:=NormalForm(g1*c,G);  
g3:=NormalForm(g2*c,G);
g4:=NormalForm(g3*b,G); 
g5:=NormalForm(g4*c,G); 
g6:=NormalForm(g5*A-(1/2)*P*x+(1/4)*P*P,G); 
\end{verbatim}
and modify the last call of the $\mathtt{printf}$ function accordingly:
\begin{verbatim}
printf"Hypersurface equation is 0 = - x^2 + "; g6; printf"\n \n";
\end{verbatim}
Finally, one can always bring the generic high-charge case into elliptic form, by duly repeating the steps explained for $D_4$. The case of $E_6$ is particularly tractable, and a simple specialization of the length-three universal flop produces e.g. the following (local) Weierstrass model \cite[Ex. 5.3]{karmazyn}:
\begin{equation}
x^2 = -y^3-\frac{3}{4}y^2T^2 + \left(\frac{3}{2}Tz^2+4T^3\right) y - \left(3T^5 +(T^2 +T^3)z^2 +\frac{1}{4}z^4\right) \ \subset \ \cc^4_{(y,x,T,z)}\ . 
\end{equation}
The explicit $\U(1)$ generator can be found via the method we explain in the next subsection. It would be interesting to see whether the charge-three (and four) models of \cite{klevers-mayorga-oehlmann-piragua-reuter,raghuram} fit into this analysis.\footnote{Notice that other examples of flops of length three have already appeared in the mathematics literature \cite{curto,curto-phd,ando}. They should all be specializations of the universal length-three flop.}

\subsubsection*{Producing matrix factorizations}

Once the deformation algebra has been input as explained in the previous section, one can also compute a matrix factorization of the universal flop of length $\ell$. To do that, one simply needs to know a set of $2\ell$ paths from the left $\cc$ vertex to the right $\cc^\ell$ one, which generate the algebra. They are given in table~\ref{tab:paths} for each $\ell$.\footnote{The table is taken from \cite[App. A.3]{karmazyn}.}
\begin{table}
\centering
\begin{tabular}{c | c | c}
length $\ell$ & $-x^2+f(z,t,T_i)$ & $2\ell$ generators \\ \hline
1 & \eqref{eq:A1verdef} & $a\ ,\ b$ \\
2 & \eqref{eq:D4verdef} & $a\ ,\ ab\ ,\ ac\ ,\ abc$ \\
3 & $-x^2 +z^2x + y^3 + \ldots$ & $a\ ,\ ac\ ,\ ac^2\ ,\ acb\ ,\ ac^2b\ ,\ ac^2bc$ \\
4 & $-x^2 + y^3 -yz^3+ \ldots$ & $a\ ,\ ac\ ,\ ac^2\ ,\ ac^3\ ,\ ac^2b\ ,\ ac^3b\ ,\ ac^3bc\ ,\ ac^3bc^2$  \\
5 & $-x^2 + y^3 +z^5+ \ldots$& $a\ ,\ ac\ ,\ ac^2\ ,\ acb\ ,\ ac^3\ ,\ ac^2b\ ,$\\ && $ac^3b\ ,\ ac^3bc\ ,\ ac^3b^2\ ,\ ac^3bc^2$ \\
6 & $-x^2 + y^3 -z^5 +\ldots$ & $a\ ,\ ac\ ,\ ac^2\ ,\ ac^2b\ ,\ ac^2bc\ ,$\\ && $ac^2bc^2\ ,\ ac^2bcb\ ,\ ac^2 bc^2b\ , ac^2bc^2bc\ ,$ \\ && $ac^2bc^2bcb\ ,\ ac^2bc^2bcbc\ ,\ ac^2bc^2bcbc^2$
\end{tabular}
\caption{The set of $2\ell$ paths $\cc \to \cc^\ell$ generating the deformation algebra. The ellipsis in the cases $\ell =3,4,5,6$ denotes the versal deformation in terms of the $T_i$.}
\label{tab:paths}
\end{table}
Then one simply modifies the code as follows:
\begin{enumerate}
\item remove $\mathtt{x}$ from the definition of $\mathtt{F}$ and decrease number of generators of the latter by one;
\item remove the relation $\mathtt{x}-\mathtt{a}\cdots\mathtt{A}$ from the definition of $\mathtt{B}$;
\item remove (or comment out) the calculation of the polynomials $\mathtt{g}_i$ (by adding $//$ in front of them); alternatively, replace $\mathtt{x}$ inside $\mathtt{g}_\text{last}$ with $\mathtt{a}\cdots\mathtt{A}$;
\item redefine the shifted $\mathtt{x}$ (i.e.\ $x-\tfrac{1}{2}P$) by adding the line
\begin{verbatim}
newx:=(a*...*A-(1/2)*P);
\end{verbatim}
\item finally, add the following piece of code:
\begin{verbatim}
L:=[a, a*b, a*c, a*b*c];

printf"The matrix factorization is C:= \n {\n "; for M in L do
printf"{ ";
for N in L do
m:=NormalForm(newx*M,G);
printf"%o", MonomialCoefficient(m,N); if N ne L[#L] then printf", ";
end if;
end for; printf"}";
if M ne L[#L] then printf", ";
end if; printf" \n ";
end for; printf"} \n \n";
\end{verbatim}
\end{enumerate}
The output is a $2\ell \times 2\ell$ matrix $C$ such that $(C,C)$ is an MF of the hypersurface with the $-x^2$ term removed, i.e. $(C+ x 1_{2\ell})(C- x 1_{2\ell})$ is an MF of $-x^2+f(z,t,T_i)$, where $f$ depends on the choice of ADE surface, and is given in table~\ref{tab:paths}. (In the above example we give the basis $\mathtt{L}$ of $2\ell=4$ paths for the $D_4$ case.)

Once an MF of the versally deformed ADE surface (i.e. universal flop of length $\ell$) is produced, one can apply the argument explained in section \ref{sub:mf} to construct a family of non-Cartier divisors.

\subsubsection*{The most general threefold models with charges three through six}

In the ancillary \texttt{Mathematica} notebook included with the \texttt{arXiv} submission,\footnote{Which can be downloaded from this \href{http://web.math.ucsb.edu/~drm/mathematica/}{\nolinkurl{page}}.} we have written the output of the outlined calculation, for the cases $E_6, E_7, E_8^{(5)}, E_8^{(6)}$ (charges three, four, five, six respectively) with all deformation parameters turned on. The outputs are in the form of a (rather lengthy) hypersurface equation directly provided in elliptic form; upon taking $z$ and the $T_i$'s to be sections of appropriate line bundles over a twofold base (satisfying relations), these should be regarded as the first examples of F-theory threefold compactifications realizing matter of charges three through six explicitly (except for the models which already appeared in \cite{klevers-mayorga-oehlmann-piragua-reuter,raghuram}).




\bibliography{mpa,v5}

\providecommand{\href}[2]{#2}\begin{thebibliography}{10}

\bibitem{enhanced}
S.~Katz, D.~R. Morrison, and M.~R. Plesser, ``Enhanced gauge symmetry in type
  {II} string theory,'' {\em Nucl. Phys. B} {\bf 477} (1996) 105--140,
  \href{http://arXiv.org/abs/arXiv:hep-th/9601108}{{\tt arXiv:hep-th/9601108}}.

\bibitem{witten-phases}
E.~Witten, ``{Phase transitions in M theory and F theory},'' {\em Nucl. Phys.}
  {\bf B471} (1996) 195--216,
\href{http://arXiv.org/abs/hep-th/9603150}{{\tt hep-th/9603150}}.

\bibitem{klemm-lerche-mayr-vafa-warner}
A.~Klemm, W.~Lerche, P.~Mayr, C.~Vafa, and N.~P. Warner, ``{Selfdual strings
  and N=2 supersymmetric field theory},'' {\em Nucl. Phys.} {\bf B477} (1996)
  746--766,
\href{http://arXiv.org/abs/hep-th/9604034}{{\tt hep-th/9604034}}.

\bibitem{katz-vafa}
S.~H. Katz and C.~Vafa, ``{Matter from geometry},'' {\em Nucl. Phys.} {\bf
  B497} (1997) 146--154,
\href{http://arXiv.org/abs/hep-th/9606086}{{\tt hep-th/9606086}}.

\bibitem{bershadsky-johansen-pantev-sadov-vafa}
M.~Bershadsky, A.~Johansen, T.~Pantev, V.~Sadov, and C.~Vafa, ``{F theory,
  geometric engineering and N=1 dualities},'' {\em Nucl. Phys.} {\bf B505}
  (1997) 153--164,
\href{http://arXiv.org/abs/hep-th/9612052}{{\tt hep-th/9612052}}.

\bibitem{katz-klemm-vafa}
S.~H. Katz, A.~Klemm, and C.~Vafa, ``{Geometric engineering of quantum field
  theories},'' {\em Nucl. Phys.} {\bf B497} (1997) 173--195,
\href{http://arXiv.org/abs/hep-th/9609239}{{\tt hep-th/9609239}}.

\bibitem{katz-mayr-vafa}
S.~Katz, P.~Mayr, and C.~Vafa, ``{Mirror symmetry and exact solution of 4-D N=2
  gauge theories: 1.},'' {\em Adv. Theor. Math. Phys.} {\bf 1} (1998) 53--114,
\href{http://arXiv.org/abs/hep-th/9706110}{{\tt hep-th/9706110}}.

\bibitem{mayr}
P.~Mayr, ``{Geometric construction of N=2 gauge theories},'' {\em Fortsch.
  Phys.} {\bf 47} (1999) 39--63,
\href{http://arXiv.org/abs/hep-th/9807096}{{\tt hep-th/9807096}}.

\bibitem{vafa}
C.~Vafa, ``{Evidence for F theory},'' {\em Nucl. Phys.} {\bf B469} (1996)
  403--418,
\href{http://arXiv.org/abs/hep-th/9602022}{{\tt hep-th/9602022}}.

\bibitem{morrison-vafa1}
D.~R. Morrison and C.~Vafa, ``{Compactifications of F theory on Calabi-Yau
  threefolds. 1},'' {\em Nucl. Phys.} {\bf B473} (1996) 74--92,
\href{http://arXiv.org/abs/hep-th/9602114}{{\tt hep-th/9602114}}.

\bibitem{morrison-vafa2}
D.~R. Morrison and C.~Vafa, ``{Compactifications of F theory on Calabi-Yau
  threefolds. 2.},'' {\em Nucl. Phys.} {\bf B476} (1996) 437--469,
\href{http://arXiv.org/abs/hep-th/9603161}{{\tt hep-th/9603161}}.

\bibitem{aspinwall-katz-morrison}
P.~S. Aspinwall, S.~H. Katz, and D.~R. Morrison, ``{Lie groups, Calabi-Yau
  threefolds, and F theory},'' {\em Adv. Theor. Math. Phys.} {\bf 4} (2000)
  95--126,
\href{http://arXiv.org/abs/hep-th/0002012}{{\tt hep-th/0002012}}.

\bibitem{aspinwall-morrison-U1}
P.~S. Aspinwall and D.~R. Morrison, ``{Nonsimply connected gauge groups and
  rational points on elliptic curves},'' {\em JHEP} {\bf 07} (1998) 012,
\href{http://arXiv.org/abs/hep-th/9805206}{{\tt hep-th/9805206}}.

\bibitem{gubser-tasi}
S.~S. Gubser, ``{TASI lectures: Special holonomy in string theory and M
  theory},'' in {\em {Strings, Branes and Extra Dimensions: TASI 2001:
  Proceedings}}, pp.~197--233.
\newblock 2002.
\newblock
\href{http://arXiv.org/abs/hep-th/0201114}{{\tt hep-th/0201114}}.
\newblock

\bibitem{Cianci:2018vwv}
F.~M. Cianci, D.~K. Mayorga~Peña, and R.~Valandro, ``{High U(1) charges in
  type IIB models and their F-theory lift},'' {\em JHEP} {\bf 04} (2019) 012,
\href{http://arXiv.org/abs/1811.11777}{{\tt 1811.11777}}.

\bibitem{klevers-mayorga-oehlmann-piragua-reuter}
D.~Klevers, D.~K. Mayorga~Pena, P.-K. Oehlmann, H.~Piragua, and J.~Reuter,
  ``{F-Theory on all Toric Hypersurface Fibrations and its Higgs Branches},''
  {\em JHEP} {\bf 01} (2015) 142,
\href{http://arXiv.org/abs/1408.4808}{{\tt 1408.4808}}.

\bibitem{raghuram}
N.~Raghuram, ``{Abelian F-theory Models with Charge-3 and Charge-4 Matter},''
  {\em JHEP} {\bf 05} (2018) 050,
\href{http://arXiv.org/abs/1711.03210}{{\tt 1711.03210}}.

\bibitem{morrison-park}
D.~R. Morrison and D.~S. Park, ``{F-Theory and the Mordell-Weil Group of
  Elliptically-Fibered Calabi-Yau Threefolds},'' {\em JHEP} {\bf 10} (2012)
  128,
\href{http://arXiv.org/abs/1208.2695}{{\tt 1208.2695}}.

\bibitem{tall-sections}
D.~R. Morrison and D.~S. Park, ``Tall sections from non-minimal
  transformations,'' {\em JHEP} {\bf 10} (2016) 033,
\href{http://arXiv.org/abs/arXiv:1606.07444 [hep-th]}{{\tt arXiv:1606.07444
  [hep-th]}}.

\bibitem{Mayrhofer:2012zy}
C.~Mayrhofer, E.~Palti, and T.~Weigand, ``{U(1) symmetries in F-theory GUTs
  with multiple sections},'' {\em JHEP} {\bf 03} (2013) 098,
\href{http://arXiv.org/abs/1211.6742}{{\tt 1211.6742}}.

\bibitem{Lawrie:2015hia}
C.~Lawrie, S.~Schafer-Nameki, and J.-M. Wong, ``{F-theory and All Things
  Rational: Surveying U(1) Symmetries with Rational Sections},'' {\em JHEP}
  {\bf 09} (2015) 144,
\href{http://arXiv.org/abs/1504.05593}{{\tt 1504.05593}}.

\bibitem{Mayrhofer:2014opa}
C.~Mayrhofer, D.~R. Morrison, O.~Till, and T.~Weigand, ``{Mordell-Weil Torsion
  and the Global Structure of Gauge Groups in F-theory},'' {\em JHEP} {\bf 10}
  (2014) 16,
\href{http://arXiv.org/abs/1405.3656}{{\tt 1405.3656}}.

\bibitem{Mayrhofer:2014haa}
C.~Mayrhofer, E.~Palti, O.~Till, and T.~Weigand, ``{Discrete Gauge Symmetries
  by Higgsing in four-dimensional F-Theory Compactifications},'' {\em JHEP}
  {\bf 12} (2014) 068,
\href{http://arXiv.org/abs/1408.6831}{{\tt 1408.6831}}.

\bibitem{taylor-turner}
W.~Taylor and A.~P. Turner, ``{An infinite swampland of U(1) charge spectra in
  6D supergravity theories},'' {\em JHEP} {\bf 06} (2018) 010,
\href{http://arXiv.org/abs/1803.04447}{{\tt 1803.04447}}.

\bibitem{raghuram-taylor}
N.~Raghuram and W.~Taylor, ``{Large U(1) charges in F-theory},'' {\em JHEP}
  {\bf 10} (2018) 182,
\href{http://arXiv.org/abs/1809.01666}{{\tt 1809.01666}}.

\bibitem{Weigand:2018rez}
T.~Weigand, ``{F-theory},'' {\em PoS} {\bf TASI2017} (2018) 016,
\href{http://arXiv.org/abs/1806.01854}{{\tt 1806.01854}}.

\bibitem{cvetic-lin-tasi}
M.~Cvetič and L.~Lin, ``{TASI Lectures on Abelian and Discrete Symmetries in
  F-theory},'' {\em PoS} {\bf TASI2017} (2018) 020,
\href{http://arXiv.org/abs/1809.00012}{{\tt 1809.00012}}.

\bibitem{whatF}
D.~R. Morrison, ``What is {F}-theory?.'' {\it To appear}.

\bibitem{klevers-taylor}
D.~Klevers and W.~Taylor, ``{Three-Index Symmetric Matter Representations of
  SU(2) in F-Theory from Non-Tate Form Weierstrass Models},'' {\em JHEP} {\bf
  06} (2016) 171,
\href{http://arXiv.org/abs/1604.01030}{{\tt 1604.01030}}.

\bibitem{klevers-morrison-raghuram-taylor}
D.~Klevers, D.~R. Morrison, N.~Raghuram, and W.~Taylor, ``{Exotic matter on
  singular divisors in F-theory},'' {\em JHEP} {\bf 11} (2017) 124,
\href{http://arXiv.org/abs/1706.08194}{{\tt 1706.08194}}.

\bibitem{kronheimer-ALE}
P.~B. Kronheimer, ``The construction of {ALE} spaces as hyper-{K}\"{a}hler
  quotients,'' {\em J. Differential Geom.} {\bf 29} (1989), no.~3, 665--683.

\bibitem{brieskorn-sim}
E.~Brieskorn, ``Singular elements of semi-simple algebraic groups,''.

\bibitem{slodowy}
P.~Slodowy, {\em Simple singularities and simple algebraic groups}, vol.~815 of
  {\em Lecture Notes in Mathematics}.
\newblock Springer, Berlin, 1980.

\bibitem{slodowy-fourlec}
P.~Slodowy, {\em Four lectures on simple groups and singularities}, vol.~11 of
  {\em Communications of the Mathematical Institute, Rijksuniversiteit
  Utrecht}.
\newblock Rijksuniversiteit Utrecht, Mathematical Institute, Utrecht, 1980.

\bibitem{katz-morrison}
S.~Katz and D.~R. Morrison, ``Gorenstein threefold singularities with small
  resolutions via invariant theory for {W}eyl groups,'' {\em J. Algebraic
  Geom.} {\bf 1} (1992), no.~3, 449--530.

\bibitem{[CKM]}
H.~Clemens, J.~Koll{\'a}r, and S.~Mori, {\em Higher Dimensional Complex
  Geometry}, vol.~166 of {\em Ast\'erisque}.
\newblock Soci{\'e}t{\'e} Math{\'e}matique de France, Paris, 1988.

\bibitem{curto-morrison}
C.~Curto and D.~R. Morrison, ``Threefold flops via matrix factorization,'' {\em
  J. Algebraic Geom.} {\bf 22} (2013), no.~4, 599--627.

\bibitem{karmazyn}
J.~{Karmazyn}, ``{The length classification of threefold flops via
  noncommutative algebras},'' {\em arXiv e-prints} (Sep, 2017)
  arXiv:1709.02720, \href{http://arXiv.org/abs/1709.02720}{{\tt 1709.02720}}.

\bibitem{stromcon}
A.~Strominger, ``Massless black holes and conifolds in string theory,'' {\em
  Nucl. Phys. B} {\bf 451} (1995) 96--108,
  \href{http://arXiv.org/abs/hep-th/9504090}{{\tt hep-th/9504090}}.

\bibitem{bhole}
B.~R. Greene, D.~R. Morrison, and A.~Strominger, ``Black hole condensation and
  the unification of string vacua,'' {\em Nucl. Phys. B} {\bf 451} (1995)
  109--120, \href{http://arXiv.org/abs/arXiv:hep-th/9504145}{{\tt
  arXiv:hep-th/9504145}}.

\bibitem{Durfee}
A.~H. Durfee, ``Fifteen characterizations of rational double points and simple
  critical points,'' {\em Enseign. Math. (2)} {\bf 25} (1979), no.~1-2,
  131--163.

\bibitem{kronheimer-torelli}
P.~B. Kronheimer, ``A {T}orelli-type theorem for gravitational instantons,''
  {\em J. Differential Geom.} {\bf 29} (1989), no.~3, 685--697.

\bibitem{remarksK3}
D.~R. Morrison, ``Some remarks on the moduli of {K3} surfaces,'' in {\em
  Classification of Algebraic and Analytic Manifolds}, K.~Ueno, ed., vol.~39 of
  {\em Progress in Math.}, pp.~303--332.
\newblock Birkh\"auser, Boston, Basel, Stuttgart, 1983.

\bibitem{MR902574}
R.~Kobayashi and A.~N. Todorov, ``Polarized period map for generalized {$K3$}
  surfaces and the moduli of {E}instein metrics,'' {\em T\^ohoku Math. J. (2)}
  {\bf 39} (1987), no.~3, 341--363.

\bibitem{MR480350}
S.~T. Yau, ``On the {R}icci curvature of a compact {K}\"ahler manifold and the
  complex {M}onge-{A}mp\`ere equation. {I},'' {\em Comm. Pure Appl. Math.} {\bf
  31} (1978), no.~3, 339--411.

\bibitem{MR1143663}
M.~T. Anderson, ``The {$L\sp 2$} structure of moduli spaces of {E}instein
  metrics on {$4$}-manifolds,'' {\em Geom. Funct. Anal.} {\bf 2} (1992), no.~1,
  29--89.

\bibitem{Hawking:1976jb}
S.~W. Hawking, ``Gravitational instantons,'' {\em Phys. Lett. A} {\bf 60}
  (1977)
81--83.

\bibitem{MR2869021}
H.-J. Hein, ``Gravitational instantons from rational elliptic surfaces,'' {\em
  J. Amer. Math. Soc.} {\bf 25} (2012), no.~2, 355--393.

\bibitem{arXiv:1505.01790}
G.~Chen and X.~Chen, ``Gravitational instantons with faster than quadratic
  curvature decay {(I)},'' \href{http://arXiv.org/abs/arXiv:1505.01790
  [math.DG]}{{\tt arXiv:1505.01790 [math.DG]}}.

\bibitem{arXiv:1508.07908}
G.~Chen and X.~Chen, ``Gravitational instantons with faster than quadratic
  curvature decay {(II)},'' \href{http://arXiv.org/abs/arXiv:1508.07908
  [math.DG]}{{\tt arXiv:1508.07908 [math.DG]}}.

\bibitem{arXiv:1603.08465}
G.~Chen and X.~Chen, ``Gravitational instantons with faster than quadratic
  curvature decay {(III)},'' \href{http://arXiv.org/abs/arXiv:1603.08465
  [math.DG]}{{\tt arXiv:1603.08465 [math.DG]}}.

\bibitem{hein-talk}
H.-J. Hein, ``{ALG and ALH spaces}.'' Seminar given at {\it Metric and Analytic
  Aspects of Moduli Spaces}, Isaac Newton Institute, University of Cambridge,
  2015.

\bibitem{Candelas:1987kf}
P.~Candelas, A.~M. Dale, C.~A. L{\"u}tken, and R.~Schimmrigk, ``Complete
  intersection {C}alabi--{Y}au manifolds,'' {\em Nuclear Phys. B} {\bf 298}
  (1988)
493--525.

\bibitem{confinement}
B.~R. Greene, D.~R. Morrison, and C.~Vafa, ``A geometric realization of
  confinement,'' {\em Nuclear Phys. B} {\bf 481} (1996) 513--538,
\href{http://arXiv.org/abs/arXiv:hep-th/9608039}{{\tt arXiv:hep-th/9608039}}.

\bibitem{clemens1983double}
C.~H. Clemens, ``Double solids,'' {\em Adv. in Math.} {\bf 47} (1983), no.~2,
  107--230.

\bibitem{MR848512}
R.~Friedman, ``Simultaneous resolution of threefold double points,'' {\em Math.
  Ann.} {\bf 274} (1986), no.~4, 671--689.

\bibitem{klebanov-witten}
I.~R. Klebanov and E.~Witten, ``{Superconformal field theory on three-branes at
  a Calabi-Yau singularity},'' {\em Nucl. Phys.} {\bf B536} (1998) 199--218,
\href{http://arXiv.org/abs/hep-th/9807080}{{\tt hep-th/9807080}}.

\bibitem{nonspherI}
D.~R. Morrison and M.~R. Plesser, ``Non-spherical horizons, {I},'' {\em Adv.
  Theor. Math. Phys.} {\bf 3} (1999) 1--81,
\href{http://arXiv.org/abs/hep-th/9810201}{{\tt hep-th/9810201}}.

\bibitem{atiyah}
M.~F. Atiyah, ``On analytic surfaces with double points,'' {\em Proc. Roy. Soc.
  London. Ser. A} {\bf 247} (1958) 237--244.

\bibitem{reid}
M.~Reid, ``Minimal models of canonical {$3$}-folds,'' in {\em Algebraic
  varieties and analytic varieties ({T}okyo, 1981)}, vol.~1 of {\em Adv. Stud.
  Pure Math.}, pp.~131--180.
\newblock North-Holland, Amsterdam, 1983.

\bibitem{unpublished}
A.~Grothendieck. {\it Unpublished}.

\bibitem{cachazo-katz-vafa}
F.~Cachazo, S.~Katz, and C.~Vafa, ``{Geometric transitions and N=1 quiver
  theories},''
\href{http://arXiv.org/abs/hep-th/0108120}{{\tt hep-th/0108120}}.

\bibitem{collinucci-fazzi-valandro}
A.~Collinucci, M.~Fazzi, and R.~Valandro, ``{Geometric engineering on flops of
  length two},'' {\em JHEP} {\bf 04} (2018) 090,
\href{http://arXiv.org/abs/1802.00813}{{\tt 1802.00813}}.

\bibitem{grimm-timo}
T.~W. Grimm and T.~Weigand, ``{On Abelian Gauge Symmetries and Proton Decay in
  Global F-theory GUTs},'' {\em Phys. Rev.} {\bf D82} (2010) 086009,
\href{http://arXiv.org/abs/1006.0226}{{\tt 1006.0226}}.

\bibitem{bcv-fate}
A.~P. Braun, A.~Collinucci, and R.~Valandro, ``{The fate of U(1)'s at strong
  coupling in F-theory},'' {\em JHEP} {\bf 07} (2014) 028,
\href{http://arXiv.org/abs/1402.4054}{{\tt 1402.4054}}.

\bibitem{aspinwall-morrison}
P.~S. Aspinwall and D.~R. Morrison, ``Quivers from matrix factorizations,''
  {\em Communications in Mathematical Physics} {\bf 313} (2012), no.~3,
  607--633.

\bibitem{FOAG}
R.~Vakil, {\em The Rising Sea: Foundations of Algebraic Geometry}.
\newblock Available at: http://math.stanford.edu/~vakil/216blog/index.html.

\bibitem{laufer}
H.~B. Laufer, ``On {$\mathbb{CP}^{1}$} as an exceptional set,'' in {\em Recent
  developments in several complex variables ({P}roc. {C}onf., {P}rinceton
  {U}niv., {P}rinceton, {N}. {J}., 1979)}, vol.~100 of {\em Ann. of Math.
  Stud.}, pp.~261--275.
\newblock Princeton Univ. Press, Princeton, N.J., 1981.

\bibitem{pinkham}
H.~C. Pinkham, ``Factorization of birational maps in dimension {$3$},'' in {\em
  Singularities, {P}art 2 ({A}rcata, {C}alif., 1981)}, vol.~40 of {\em Proc.
  Sympos. Pure Math.}, pp.~343--371.
\newblock Amer. Math. Soc., Providence, RI, 1983.

\bibitem{morrison}
D.~R. Morrison, ``The birational geometry of surfaces with rational double
  points,'' {\em Math. Ann.} {\bf 271} (1985), no.~3, 415--438.

\bibitem{Rossi:2004eq}
M.~Rossi, ``{Geometric transitions},'' {\em J. Geom. Phys.} {\bf 56} (2006)
  1940--1983,
\href{http://arXiv.org/abs/math/0412514}{{\tt math/0412514}}.

\bibitem{Acharya:2004qe}
B.~S. Acharya and S.~Gukov, ``{M theory and singularities of exceptional
  holonomy manifolds},'' {\em Phys. Rept.} {\bf 392} (2004) 121--189,
\href{http://arXiv.org/abs/hep-th/0409191}{{\tt hep-th/0409191}}.

\bibitem{Candelas:1989ug}
P.~Candelas, P.~S. Green, and T.~Hubsch, ``{Rolling Among Calabi-Yau Vacua},''
  {\em Nucl. Phys.} {\bf B330} (1990)
49.

\bibitem{Strominger:1995cz}
A.~Strominger, ``{Massless black holes and conifolds in string theory},'' {\em
  Nucl. Phys.} {\bf B451} (1995) 96--108,
\href{http://arXiv.org/abs/hep-th/9504090}{{\tt hep-th/9504090}}.

\bibitem{Greene:1995hu}
B.~R. Greene, D.~R. Morrison, and A.~Strominger, ``{Black hole condensation and
  the unification of string vacua},'' {\em Nucl. Phys.} {\bf B451} (1995)
  109--120,
\href{http://arXiv.org/abs/hep-th/9504145}{{\tt hep-th/9504145}}.

\bibitem{magma}
W.~Bosma, J.~Cannon, and C.~Playoust, ``The {M}agma algebra system. {I}. {T}he
  user language,'' {\em J. Symbolic Comput.} {\bf 24} (1997), no.~3-4,
  235--265. Computational algebra and number theory (London, 1993).

\bibitem{curto}
C.~Curto, ``{Matrix model superpotentials and ADE singularities},'' {\em Adv.
  Theor. Math. Phys.} {\bf 12} (2008), no.~2, 353--404,
\href{http://arXiv.org/abs/hep-th/0612172}{{\tt hep-th/0612172}}.

\bibitem{curto-phd}
C.~Curto, {\em Matrix model superpotentials and {C}alabi-{Y}au spaces: {A}n
  {A}-{D}-{E} classification}.
\newblock ProQuest LLC, Ann Arbor, MI, 2005.
\newblock Thesis (Ph.D.)--Duke University.

\bibitem{ando}
T.~Ando, ``Some examples of simple small singularities,'' {\em Comm. Algebra}
  {\bf 41} (2013), no.~6, 2193--2204.

\end{thebibliography}
\bibliographystyle{at}

\end{document}